\documentclass[pra, twocolumn, letterpaper,superscriptaddress]{revtex4}
\usepackage{amsmath}
\usepackage{amssymb}
\usepackage{mathrsfs}
\usepackage[mathscr]{euscript}	%For nice fonts with \mathscr
\usepackage{xspace}
\usepackage{graphicx}				%Figures
\usepackage{braket}					%Dirac braket notation
\usepackage[utf8]{inputenc}
\usepackage{enumitem}			%for itemized with no margins
\usepackage{tensor}

\usepackage{easybmat}

\setcitestyle{numbers,square}
\usepackage{hyperref}
\hypersetup{
  colorlinks   	= true, 	%Colours links instead of ugly boxes
  urlcolor     	= blue, 	%Colour for external hyperlinks
  linkcolor    	= blue, 	%Colour of internal links
  citecolor   	= blue 	%Colour of citations
}
\usepackage{ifpdf}
\ifpdf
\fi

\newcommand{\eq}[1]{(\ref{#1})}
\newcommand{\Eq}[1]{Eq.~(\ref{#1})}
\newcommand{\Eqs}[1]{Eqs.~(\ref{#1})}
\newcommand{\Fig}[1]{Fig.~\ref{#1}}
\newcommand{\Sec}[1]{Sec.~\ref{#1}}
\newcommand{\Ref}[1]{Ref.~\cite{#1}}
\newcommand{\Refs}[1]{Refs.~\cite{#1}}
\newcommand{\App}[1]{Appendix~\ref{#1}}

\newcommand{\eg}{{e.g.,\/}\xspace}
\newcommand{\ie}{{i.e.,\/}\xspace}

\newcommand{\pd}{\partial}
\newcommand{\del }{\vec{\nabla}}
\newcommand{\cc}{\text{c.\,c.}}
\newcommand{\hc}{\text{h.\,c.}}

\newcommand{\mc}[1]{\mathcal{#1}}

\newcommand{\mcu}[1]{\mathscr{#1}}
\newcommand{\oper}[1]{\hat{\mcu{#1}} }
\renewcommand{\vec}[1]{{\boldsymbol{\rm #1}}}

\begin{document}

\title{Extending geometrical optics: A Lagrangian theory for vector waves}

\begin{abstract}

%Even when neglecting diffraction effects, the well-known equations of geometrical optics (GO) are not entirely accurate. Traditional GO treats wave rays as classical particles, which are completely described by their coordinates and momenta, but vector wave rays have another degree of freedom, namely, their polarization. The polarization degree of freedom manifests itself as an effective (classical) spin that can be assigned to rays and can affect the wave dynamics accordingly. Manifestations of wave-spin dynamics include mode conversion and polarization-driven bending of ray trajectories. This work presents an extension and reformulation of GO as a first-principle Lagrangian theory, whose effective-gauge Hamiltonian governs the aforementioned polarization phenomena simultaneously. As an example, the theory is applied to describe the polarization-driven divergence of right-hand and left-hand circularly polarized electromagnetic waves in weakly magnetized plasma.

Even when neglecting diffraction effects, the well-known equations of geometrical optics (GO) are not entirely accurate. Traditional GO treats wave rays as classical particles, which are completely described by their coordinates and momenta, but vector-wave rays have another degree of freedom, namely, their polarization. The polarization degree of freedom manifests itself as an effective (classical) ``wave spin" that can be assigned to rays and can affect the wave dynamics accordingly. A well-known manifestation of polarization dynamics is mode conversion, which is the linear exchange of quanta between different wave modes and can be interpreted as a rotation of the wave spin. Another, less-known polarization effect is the polarization-driven bending of ray trajectories. This work presents an extension and reformulation of GO as a first-principle Lagrangian theory, whose effective-gauge Hamiltonian governs the aforementioned polarization phenomena simultaneously. As an example, the theory is applied to describe the polarization-driven divergence of right-hand and left-hand circularly polarized electromagnetic waves in weakly magnetized plasma.

\end{abstract}

\author{D.~E. Ruiz}
\affiliation{Department of Astrophysical Sciences, Princeton University, Princeton, New Jersey 08544, USA}
\author{I.~Y. Dodin}
\affiliation{Department of Astrophysical Sciences, Princeton University, Princeton, New Jersey 08544, USA}
\affiliation{Princeton Plasma Physics Laboratory, Princeton, New Jersey 08543, USA}

\date{\today}

\maketitle

%%%%%%%%%%%%%%%%%%%%%%%%%%%%%%%%%%%%%%%
%%%%%%%%%%%%%%%%%%%%%%%%%%%%%%%%%%%%%%%
%%%%%%%%%%%%%%%%%%%%%%%%%%%%%%%%%%%%%%%
\section{Introduction}

%%%%%%%%%%%%%%%%%%%%%%%%%%%%%%%%%%%%%%%
\subsection{Motivation}

Geometrical optics (GO) is a reduced model of wave dynamics \cite{Tracy:2014to,Whitham:2011kb} that is widely used in many contexts ranging from quantum dynamics to electromagnetic (EM), acoustic, and gravitational phenomena \cite{Whitham:1965kx,Dodin:2012hn,Isaacson:1968vm}. Mathematically, GO is an asymptotic theory with respect to a small parameter $\epsilon$ that is a ratio of the wave relevant characteristic period (temporal or spatial) to the inhomogeneity scale of the underlying medium. Practical applications of GO are traditionally restricted to the lowest-order theory, where each wave is basically approximated with a local eigenmode of the underlying medium at each given spacetime location. Then, the wave dynamics is entirely determined by a single branch of the local dispersion relation. However, this approximation is not entirely accurate, even when diffraction is neglected. If a dispersion relation has more than one branch, \ie a vector wave with more than one polarization at a given location, then the interaction between these branches can give rise to important polarization effects that are missed in the traditional lowest-order GO.

One interesting manifestation of such polarization effects is the polarization-driven bending of ray trajectories. At the present moment, it is known primarily in two contexts. One is quantum mechanics, where polarization effects manifest as the Berry phase \cite{Berry:1984jv} and the associated Stern-Gerlach force experienced by vector particles, \ie quantum particles with spin. Another one is optics, where a related effect has been known as the \emph{Hall effect of light}; namely, even in an isotropic dielectric, rays propagate somewhat differently depending on polarization if the dielectric is inhomogeneous (see, c.f.  \Refs{Bliokh:2008km,Bliokh:2015bw,Onoda:2004ij,Dooghin:1992kr,Liberman:1992bz}). But the same effect can also be anticipated for waves in plasmas, \eg radiofrequency (RF) waves in tokamaks. In fact, since $\epsilon$ for RF waves in laboratory plasma is typically larger than that for quantum and optical waves, the polarization-driven bending of ray trajectories in this case can be more important and perhaps should be taken into account in practical ray-tracing simulations. However, \textit{ad~hoc} theories of polarization effects available from optics are inapplicable to plasma waves, which have more complicated dispersion and thus require more fundamental approaches. Thus, a different theory is needed that would allow the calculation of the polarization-bending of the ray trajectories for plasma waves and, also more broadly, waves in general linear media.

Relevant work was done in \Refs{Littlejohn:1991jv,Weigert:1993vy}, where a systematic procedure was proposed to asymptotically diagonalize the dispersion operator for linear vector waves. Polarization effects emerge as $\mc{O}(\epsilon)$ corrections to the GO dispersion relation. However, this approach excludes mode conversion, \ie the linear exchange of quanta between different branches of the local dispersion relation. Since the group velocities of the different branches eventually separate, mode conversion is typically followed by ray splitting and, in this particular context, was studied extensively (see, \eg  \Refs{Friedland:1985ko,Friedland:1987um,Tracy:1993el,Flynn:1994cn,Tracy:2003bl, Tracy:2007jp, Richardson:2008fl}). However, these works considered wave modes that are resonant in small, localized regions of phase space. Hence, the nonadiabatic dynamics was formulated as an asymptotic scattering problem between two wave modes, so the polarization-driven bending of ray trajectories was not included.

The main message of this work is that mode conversion and the polarization-driven bending of ray trajectories are two sides of the same coin and can be considered simultaneously within a unified theory. The first general theory that captures them both simultaneously was proposed in \Ref{Ruiz:2015hq}. This theory was successfully benchmarked \cite{Ruiz:2015dv} against previous theories describing the \textit{Hall effect of light} \cite{Bliokh:2008km,Bliokh:2015bw,Onoda:2004ij,Dooghin:1992kr,Liberman:1992bz}. However, the formulation in \Ref{Ruiz:2015hq} is still limited since it requires that the wave equation be brought to a certain (multisymplectic) form resembling the Dirac equation. Although any nondissipative vector wave allows for such representation in principle \cite{Dodin:2014hw,Ruiz:2015hq}, casting the wave dynamics into the specific framework adopted in \Ref{Ruiz:2015hq} can be complicated. Thus, practical applications require a more flexible formulation that do not rely on this specific framework.

Here we propose such a theory. In addition to generalizing the results of \Ref{Ruiz:2015hq}, we also introduce, in a unified context and an instructive manner, some of the related advances that were made recently in \Refs{Ruiz:2015dh,Ruiz:2015bz,Dodin:2016ut}. It is expected that the comprehensive analysis presented in this work will facilitate future practical implementations of the proposed theory, particularly in improving ray-tracing simulations.

%%%%%%%%%%%%%%%%%%%%%%%%%%%%%%%%%%%%%%%
\subsection{Outline}

We consider general linear nondissipative waves determined by some Hermitian dispersion operator. Using the Feynman reparameterization and the Weyl calculus, we obtain a reduced Lagrangian for such waves. In contrast with the traditional GO Lagrangian, which has an accuracy of $\mc{O}(\epsilon^0)$ in the GO parameter $\epsilon$, our Lagrangian is $\mc{O}(\epsilon^1)$-accurate, so it captures polarization effects. As an example, we apply the formulation to study polarization effects on the propagation of EM waves propagating in weakly magnetized plasma. (The case of strongly magnetized plasma will be discussed in a separate paper.)

The advantages of our theory are as follows. (i)~The theory is derived in a variational form, so the resulting equations are manifestly conservative. (ii)~Through the use of the Feymann reparameterization, we can obtain the dynamics of continuous waves and of their rays directly from a variational principle. (iii)~The theory assumes no specific wave equation, so quantum spin effects and classical polarization effects can be studied on the same footing. (iv)~Moreover, a related formalism \cite{Ruiz:2015hq} is applicable to develop new reduced theories for relativistic spinning particles \cite{Ruiz:2015hq,Ruiz:2015dh}.

This paper is organized as follows. In \Sec{sec:notation}, the basic notation is defined. In \Sec{sec:basic}, the variational formalism used to describe vector waves is presented. In \Sec{sec:eigen}, a general procedure is proposed to block-diagonalize the wave dispersion operator. In \Sec{sec:reduced}, we reparametrize the wave action to facilitate asymptotic analysis. In \Sec{sec:GO}, the leading order GO approximation is discussed. In \Sec{sec:XGO}, the more accurate model that includes polarization effects is discussed. In \Sec{sec:discussion}, the theory is applied to describe polarization effects on the propagation of EM waves in weakly magnetized plasma. In \Sec{sec:conclusions}, our main results are summarized. Finally, \App{app:Weyl} presents a brief introduction to the Weyl symbol calculus.

%%%%%%%%%%%%%%%%%%%%%%%%%%%%%%%%%%%%%%%
%%%%%%%%%%%%%%%%%%%%%%%%%%%%%%%%%%%%%%%
%%%%%%%%%%%%%%%%%%%%%%%%%%%%%%%%%%%%%%%
\section{Notation}
\label{sec:notation}

The following notation is used throughout the paper. The symbol ``$\doteq$'' denotes definitions, ``\cc '' denotes ``complex conjugate,'' and ``\hc'' denotes ``Hermitian conjugate." The identity $N\times N$ matrix is denoted by $\mathbb{I}_{N}$. The Minkowski metric is adopted with signature $(+, -, -, -)$. Greek indices span from $0$ to $3$ and refer to spacetime coordinates $x^\mu=(x^0, \vec{x})$ with $x^0$ corresponding to the time variable $t$. Also, partial derivatives on spacetime will be denoted by $\pd_{x}$, where the individual components are $\pd_\mu \doteq \pd/\pd x^\mu=(\pd_t, \del)$ and $\mathrm{d}^4 x \doteq \mathrm{d}t\,\mathrm{d}^3\vec{x}$. Latin indices span from $1$ to $3$ and denote the spatial variables, \ie $\vec{x} = (x^1, x^2, x^3)$ and $\pd_i \doteq \pd/\pd x^i$. Summation over repeated indexes is assumed. In particular, for arbitrary four-vectors $a$ and $b$, we have $a \cdot b \doteq a^\mu b_\mu = a^0 b^0 - \vec{a}\cdot \vec{b}$. In Euler-Lagrange equations (ELEs), the notation ``$\delta a:$'' means that the corresponding equation is obtained by extremizing the action integral with respect to $a$.

%%%%%%%%%%%%%%%%%%%%%%%%%%%%%%%%%%%%%%%
%%%%%%%%%%%%%%%%%%%%%%%%%%%%%%%%%%%%%%%
%%%%%%%%%%%%%%%%%%%%%%%%%%%%%%%%%%%%%%%
\section{Basic Equations}
\label{sec:basic}

%%%%%%%%%%%%%%%%%%%%%%%%%%%%%%%%%%%%%%%
\subsection{Wave action principle}
\label{sec:WAP}

The dynamics of any nondissipative linear wave can be described by the least action principle, $\delta \mc{S} =0$, where the real action $\mc{S}$ is bilinear in the wave field \cite{Dodin:2014hw}. We represent a wave field, either classical or quantum, as a complex-valued vector $\Psi(x)$. We allow this vector field to have an arbitrary number of components $\bar{N}$. In the absence of parametric resonances \cite{foot:parametric}, the action can be written in the form \cite{Kaufman:1987ch}
\begin{equation}
	\mc{S}	\doteq	\int \mathrm{d}^4 x \, \mathrm{d}^4x' \, \Psi^\dag (x)  \mcu{D}(x,x') \Psi(x'),
	\label{def:action_x}
\end{equation}
where $\mcu{D}$ is a $\bar{N} \times \bar{N}$ Hermitian  matrix kernel $[\mcu{D}(x,x')=\mcu{D}^\dag(x',x)]$ that describes the underlying medium. Varying $\mc{S}$ with respect to $\Psi^\dag$ leads to
\begin{equation}
		\delta \Psi^\dag: 	\quad	
							0=\int 		\mathrm{d}^4 x' \,	\mcu{D}(x,x') \Psi(x') .
		\label{eq:basic_ELE}
\end{equation}
Similarly, varying with respect to $\Psi$ gives the equation adjoint to \Eq{eq:basic_ELE}, which we do not need to discuss.

It is convenient to describe the wave $\Psi(x)$ as an abstract vector $\ket{\Psi}$ in the Hilbert space of wave states with inner product \cite{Dodin:2014hw,Littlejohn:1993bd}
\begin{equation}
	\braket{ \Upsilon | \Psi } \doteq \int \mathrm{d}^4 x \, \Upsilon^\dag(x) \Psi(x). 
\end{equation}
In this representation, $\Psi(x)= \braket{x | \Psi}$, where $\ket{x}$ are the eigenstates of the coordinate operator $\hat{x}$ such that $\smash{\braket{ x | \hat{x}^\mu | x' } = x^\mu \braket{ x  | x' } = x^\mu \delta^4(x-x')}$. We also introduce the momentum (wavevector) operator $\hat{p}$ such that $\smash{\braket{ x | \hat{p}_\mu | x' } = i \pd [ \delta^4(x-x')] / \pd x^\mu }$ in the $x$-representation \cite{foot:momentum}. Thus, the action \eq{def:action_x} can be rewritten as
\begin{equation}
	\mc{S}	=	\braket{\Psi | \oper{D} | \Psi},
	\label{def:action}
\end{equation}
where $\oper{D}$ is the Hermitian \textit{dispersion operator} such that $\smash{ \mcu{D}(x,x') = \braket{ x | \oper{D} | x'} }$. Treating $\bra{\Psi}$ and $\ket{\Psi}$ as independent variables \cite{Dodin:2014hw} and varying the action \eq{def:action} gives
\begin{equation}
		\delta \bra{\Psi}: 	\quad		\oper{D} \ket{\Psi} =0	
		\label{eq:ELE_abs}		
\end{equation}
(plus the conjugate equation), which is the generalized vector form of \Eq{eq:basic_ELE}. Specifically, \Eq{eq:basic_ELE} is obtained by projecting \Eq{eq:ELE_abs} with $\bra{x}$ and using the fact that the operator $\int \mathrm{d}^4 x \, \ket{x}\bra{x} = \hat{1}$ is an identity operator.

%%%%%%%%%%%%%%%%%%%%%%%%%%%%%%%%%%%%%%%
\subsection{Extended wave function}

As shown in \Refs{Dodin:2014hw,Ruiz:2015hq}, reduced models of wave propagation are convenient to develop when the action is of the symplectic form; namely,
\begin{equation}
	\mc{S}_{\rm symplectic}	\doteq	
			\braket{\Psi | ( \hat{p}_0 \mathbb{I}_{\bar{N}} - \oper{H} ) | \Psi},
	\label{def:act_symplectic}
\end{equation}
where $\hat{p}_0 = i \pd_t$ (in the $x$-representation) and ``the wave Hamiltonian'' $\oper{H}= \mcu{H} (\hat{x}, \hat{\vec{p}})$ is some Hermitian operator that is local in time, \ie commutes with $\hat{t}$. (For extended discussions, see \Refs{Dodin:2014hw,Bridges:2001aa}.)

In order to cast the general action \eq{def:action} into the symplectic form \eq{def:act_symplectic}, let us perform the so-called Feynman reparameterization \cite{Feynman:1951ic,Aparicio:1995eh} that lifts the wave dynamics governed by \Eq{def:action} from $\mathbb{R}^4$ to $\mathbb{R}^5$. Specifically, we let the wave field depend on spacetime and on some parameter $\tau$ so that $\Psi(\tau,x) = \braket{x | \Psi (\tau) } $. Note that $\ket{\Psi (\tau)}$ belongs to the same Hilbert space defined in \Sec{sec:WAP}. Thus, the inner product remains the same; \ie $\braket{ \Upsilon (\tau') | \Psi (\tau)} = \int \mathrm{d}^4 x \, \Upsilon^\dag(\tau',x) \Psi(\tau,x)$. We consider the following ``extended'' action:
\begin{equation}
	\mc{S}_{\rm X} \doteq \int \mathrm{d} \tau ~ L,
	\label{eq:action}
\end{equation}
where $L \doteq L_\tau + L_D$,
\begin{subequations}	\label{eq:Lagragians_aux}
	\begin{gather}
		 L_{\tau}  \doteq  - (i/2)
		 						\left[ \, \braket{\Psi (\tau)| \pd_\tau  \Psi (\tau)} 
		 								- \cc	\, \right],
					\label{eq:L1_orig} \\
		 L_D  \doteq  \braket{ \Psi (\tau) | \oper{D} | \Psi (\tau)},
					\label{eq:L2_orig}
	\end{gather}
\end{subequations}
and $\pd_\tau \Psi(\tau,x) = \braket{x |  \pd_\tau  \Psi (\tau)}$. Note that the Lagrangian $L$ is local in the parameter $\tau$; \ie the abstract vectors are all evaluated at $\tau$. From hereon, all fields will be evaluated at $\tau$, and we will avoid mentioning the dependence of $\ket{\Psi}$ on $\tau$ explicitly. The ELE corresponding to the action $\mc{S}_{\rm X}$ is given by
\begin{equation}
	i \pd_\tau \ket{ \Psi} = \oper{D} \ket{ \Psi}.
	\label{eq:orig_ext}
\end{equation}
Note that \Eq{eq:orig_ext} can be interpreted as a vector Schrödinger equation in the extended variable space, where $\smash{\oper{D}}$ acts as the Hamiltonian operator. The dynamics of the original system described by \Eq{eq:ELE_abs} is a special case of the dynamics governed by \Eq{eq:orig_ext}, which corresponds to a steady state with respect to the parameter $\tau$; \ie $\pd_\tau \Psi = 0$. The advantage of the representation \eq{eq:action} is that the action has the manifestly symplectic form, so we can proceed as follows.

%%%%%%%%%%%%%%%%%%%%%%%%%%%%%%%%%%%%%%%
%%%%%%%%%%%%%%%%%%%%%%%%%%%%%%%%%%%%%%%
%%%%%%%%%%%%%%%%%%%%%%%%%%%%%%%%%%%%%%%
\section{Eigenmode representation}
\label{sec:eigen}

%%%%%%%%%%%%%%%%%%%%%%%%%%%%%%%%%%%%%%%
\subsection{Variable transformation}

We introduce a unitary $\tau$-independent transformation $\smash{\oper{Q}}$ that maps $\ket{\Psi}$ to some $\bar{N}$-dimensional abstract vector $\ket{\bar{\psi}}$ yet to be defined:
\begin{gather}
	\ket{\Psi }		 = \oper{Q} \ket{\bar{\psi}} .
	\label{eq:trans}
\end{gather}
Inserting \Eq{eq:trans} into \Eqs{eq:Lagragians_aux} leads to
\begin{subequations}
	\begin{gather}
		 L_\tau 	= - (i/2) \left( \, \braket{ \bar{\psi} | \pd_\tau \bar{\psi} } 
		 								- \cc	\, \right),
					\label{eq:L1_trans} \\
		 L_D 	=   \braket{ \bar{\psi} | \oper{D}_{\rm eff} | \bar{\psi} },
					\label{eq:L2_trans}
	\end{gather}
\end{subequations}
where $\smash{ \oper{D}_{\rm eff} \doteq \oper{Q}^\dag \oper{D} \oper{Q} }$. In what follows, we seek to construct $\smash{\oper{Q}}$ such that the operator $\smash{ \oper{D}_{\rm eff} }$ is expressed in a block-diagonal form. The procedure used is identical to that given in \Refs{Littlejohn:1991jv,Weigert:1993vy}. However, in order to account for resonant-mode coupling, $\smash{ \oper{D}_{\rm eff}}$ will be made only block-diagonal, instead of fully diagonal as in \Refs{Littlejohn:1991jv,Weigert:1993vy}.

%%%%%%%%%%%%%%%%%%%%%%%%%%%%%%%%%%%%%%%
\subsection{Weyl representation}

Let us consider \Eq{eq:L2_trans} in the Weyl representation. (Readers who are not familiar with the Weyl calculus are encouraged to read \App{app:Weyl} before continuing further.) In this representation, $L_D$ is written as \cite{Kaufman:1987ch}
\begin{equation}
	L_D =   \mathrm{Tr}  \int \mathrm{d}^4 x \, \mathrm{d}^4p \, D_{\rm eff}(x,p) W(\tau,x,p), 
	\label{eq:action_eff}
\end{equation}
where `Tr' represents the matrix trace. The Wigner tensor $W(\tau,x,p)$ corresponding to $\ket{\bar{\psi}}$ is defined as
\begin{equation}
	\tensor{W}{^m_n}(\tau,x,p) \doteq \int \frac{\mathrm{d}^4 s}{(2\pi )^4} \, 
					e^{i  p \cdot s  } 
					\braket{x+ \frac{s}{2} | \bar{\psi}^m} 
					\braket{ \bar{\psi}_n | x- \frac{s}{2} } ,
	\label{def:wigner_function}
\end{equation}
and $D_{\mathrm{eff}}( x,p)$ is the Weyl symbol [\Eq{def:weyl_symbol}] corresponding to the operator $\oper{D}_{\rm eff}$. It can be written explicitly as
\begin{equation}
	D_{\mathrm{eff}}( x,p) = [Q^\dag] (x,p) \star D(x,p) \star Q(x,p),
	\label{eq:D_eff}
\end{equation}
where `$\star$' is the Moyal product [\Eq{def:Moyal}] and $D(x,p)$, $Q(x,p)$, and $[Q^\dag](x,p)$ are the Weyl symbols corresponding to $\smash{\oper{D}}$, $\smash{\oper{Q}}$, and $\smash{\oper{Q}^\dag}$, respectively. Also, the Weyl representation of the unitary condition, $\smash{\hat{Q}^\dag \hat{Q} = \hat{\mathbb{I}}_{\bar{N}}}$, is
\begin{equation}
	[Q^\dag] (x,p) \star Q(x,p) = \mathbb{I}_{\bar{N} },
	\label{eq:T_unit}
\end{equation}
which will be used below.

%%%%%%%%%%%%%%%%%%%%%%%%%%%%%%%%%%%%%%%
\subsection{Eigenmode representation}
\label{sec:eigenmode}

Let us assume that the symbols $D_{\rm eff}$ and $Q$ can be expanded in powers of the GO parameter
\begin{equation}
	\epsilon = \mathrm{max} 
					\left\{ \frac{1}{\omega T }, \frac{1}{ |\vec{k}| \ell } \right\} \ll 1,
\end{equation}
where $\omega$ and $|\vec{k}|$ are understood as the characteristic wave frequency and wave number, respectively. Also, $T$ and $\ell$ are the characteristic time and length scales of the background medium, correspondingly. Hence, we write
\begin{subequations}
	\begin{gather}
		D_{\rm eff}(x,p)  = \Lambda(x,p) 
						+  \epsilon U(x,p)+ \mc{O}(\epsilon^2) ,
						\label{eq:series} \\
		Q(x,p) 	= Q_0(x,p)
						+  \epsilon Q_1(x,p)+ \mc{O}(\epsilon^2),
						\label{eq:series_T}
	\end{gather}
\end{subequations}
where $(\Lambda, U, Q_0, Q_1)$ are $\bar{N} \times \bar{N}$ matrices of order unity.

To the lowest-order in $\epsilon$, the Moyal products in \Eqs{eq:D_eff} and \eq{eq:T_unit} reduce to ordinary products, so
\begin{gather}
	\Lambda(x,p) = [Q_0^\dag] (x, p)  D(x, p)  Q_0  (x, p), \label{eq:lambda_equation}\\
	[Q_0^\dag] (x,p) Q_0(x,p) = \mathbb{I}_{\bar{N}}		\label{eq:unit_0}		.
\end{gather}
By properties of the Weyl transformation, the fact that $\oper{D}$ is a Hermitian operator ensures that $D(x, p)$ is a Hermitian matrix. Hence, $D(x,p)$ has $\bar{N}$ orthonormal eigenvectors $\vec{e}_q(x,p)$, which correspond to some real eigenvalues $\lambda^{(q)}(x,p)$. Let us construct $Q_0(x, p)$ out of these eigenvectors so that
\begin{equation}
	Q_0  (x, p) =
		\begin{bmatrix}
			\vec{e}_1(x,p), \, ... \,  , \, \vec{e}_{\bar{N}} (x,p)
		\end{bmatrix},
	\label{eq:T0}
\end{equation} 
where the individual $\vec{e}_q$ form the columns of $Q_0$. From \Eq{eq:unit_0}, we then find $\smash{ [Q_0^\dag] (x,p) = Q_0^\dag (x,p) }$. Hence, the matrix $\Lambda$ has the following diagonal form:
\begin{equation}
	\Lambda( x,p ) = \mathrm{diag} \,	
				[ \, \lambda^{(1)}(x,p), \, ...  \, , \, \lambda^{(\bar{N})}(x,p) \,  ].
	\label{eq:Lambda}
\end{equation}

To the next order in $\epsilon$, \Eq{eq:T_unit} reads as follows:
\begin{equation}
	Q_0^\dag Q_1 + [Q_1^\dag ] Q_0 + (i/2) \{ Q_0^\dag, Q_0 \} =0.
	\label{eq:norm_1}
\end{equation}
Here we assumed that term that involves the Poisson bracket $\{ Q_0^\dag, Q_0 \}$, which arises from the expansion of the Moyal star product [\Eq{eq:Moyal_exp}], is of the first order in $\epsilon$. Following \Ref{Littlejohn:1991jv}, we let $Q_1 = Q_0 (A+i G) $ and $\smash{ [Q_1^\dag] = Q_1^\dag }$, where $A (x,p)$ and $G (x , p )$ are $\bar{N} \times \bar{N}$ Hermitian matrices. Then, \Eq{eq:norm_1} gives
\begin{equation}
	A(x , p ) = - (i/4) \{ Q_0^\dag, Q_0 \}. 
\end{equation}
In order to determine $G(x,p)$, we write \Eq{eq:D_eff} to the first order in $\epsilon$. Introducing the bracket
\begin{equation}
	\left\lbrace A, B \right\rbrace_C 
		\doteq	\frac{\pd A}{\pd p_\mu} C \frac{\pd B}{\pd x^\mu} 
					- \frac{\pd A}{\pd x^\mu} C \frac{\pd B}{\pd p_\mu}
	\label{eq:poisson2}
\end{equation}
and noting that  $ D   Q_0 = Q_0 \Lambda $, we obtain \cite{Littlejohn:1991jv}
\begin{widetext}
	\begin{align}
	U(x,p) 	= 		&\, Q_1^\dag D Q_0 + Q_0^\dag D Q_1 
								+ (i/2)\{ Q_0^\dag  D, Q_0 \} 
								+(i/2)\{ Q_0^\dag , D  \} Q_0 
								\notag \\
				=		&	\,	(A-iG) Q_0^\dag D Q_0
								+ Q_0^\dag D Q_0 (A+iG) 
								+ (i/2)\{  Q_0^\dag  D, Q_0 \} 
								+(i/2)\{ Q_0^\dag , D  \} Q_0 
								\notag \\
				=		&	\,	A \Lambda  -i G \Lambda 
								 + \Lambda A + i \Lambda G
								+ (i/2)\{ Q_0^\dag  D, Q_0 \} 
								+(i/2)\{ Q_0^\dag  ,  D Q_0 \} 
								- (i/2)\{ Q_0^\dag  ,  Q_0 \}_D
								\notag \\
				=		&	\,	i \Lambda G -i G \Lambda 
								- (i/4)  \{ Q_0^\dag, Q_0 \} \Lambda
								- (i/4) \Lambda \{ Q_0^\dag, Q_0 \} 
								+ (i/2)\{  \Lambda Q_0^\dag, Q_0 \}  
								+(i/2)\{ Q_0^\dag  ,  Q_0 \Lambda \} 
								- (i/2)\{ Q_0^\dag  , Q_0 \}_D
								\notag \\
				=		&	\,	i ( \Lambda G - G \Lambda 
								+ \delta U ) ,
	\label{eq:G}
	\end{align}
where $\delta U(x,p)$ is a $\bar{N} \times \bar{N}$ matrix given by
\begin{equation}
	 \delta U(x,p) = (1/4)  \{ Q_0^\dag, Q_0 \} \Lambda
								+ (1/4) \Lambda \{ Q_0^\dag, Q_0 \} 
								+ (1/2)\{  \Lambda,  Q_0 \}_{Q_0^\dag} 
								+(1/2)\{ Q_0^\dag  ,   \Lambda \}_{Q_0} 
								- (1/2)\{ Q_0^\dag  ,Q_0 \}_D .
\end{equation}
\end{widetext}

Since $\tensor{(\Lambda G - G \Lambda)}{^m_n}= \tensor{G}{^m_n} [ \lambda^{(m)} - \lambda^{(n)} ]$ (no summation is assumed here over the repeating indices), one can diagonalize $U$ by adopting $\tensor{G}{^m_n} =  \tensor{\delta U}{^m_n}/[ \lambda^{(n)} - \lambda^{(m)} ]$, as done in \Refs{Littlejohn:1991jv,Weigert:1993vy}. However, this method is applicable only when $|  \lambda^{(m)} - \lambda^{(n)} | \gtrsim \mc{O}(1)$; otherwise, when $| \lambda^{(m)} - \lambda^{(n)} | \simeq \mc{O}(\epsilon)$, $G \gtrsim \mc{O}(\epsilon^{-1})$ so $\epsilon Q_1 \gtrsim \mc{O}(1)$, which is in violation of the assumed ordering in \Eq{eq:series_T}. Hence, instead of diagonalizing $U$, we propose to only \textit{block-diagonalize} $U$ as follows. When $| \lambda^{(m)} - \lambda^{(n)} | \gtrsim \mc{O}(1)$, we choose the off-components of $\tensor{G}{^m_n}$ so that $\tensor{U}{^m_n} =0$. (We call such modes \emph{nonresonant}.) When $| \lambda^{(m)} - \lambda^{(n)} | \simeq \mc{O}(\epsilon)$, we let $\tensor{G}{^m_n}=0$. (We call such modes \emph{resonant}.) By following this prescription and permutating the matrix rows, we obtain $U$ in the following form:
\begin{equation}
	U(x,p) =
	\begin{pmatrix}
		U_{1}(x,p) 		&		0					& 		\hdots		&		0		\\
		0					&		U_{2}(x,p)		&		\hdots		&		0		\\
		\vdots			&		\vdots					&		\ddots		&		\vdots 		\\
		0					&		0							&		0				&		U_J(x,p) 		
	\end{pmatrix},
	\label{eq:block_U}
\end{equation}
where $U_{j} (x,p) $ are $n_{j} \times n_{j}$ Hermitian matrices and $J$ is the total number of blocks, so $\smash{ \sum_{ j =	1 }^J n_j = \bar{N} }$. Note that, in the particular case where only nonresonant modes are present, $U(x,p)$ is diagonal, and one recovers the results obtained in \Refs{Littlejohn:1991jv,Weigert:1993vy}.

Since the matrix $D_{\rm eff} \approx \Lambda + \epsilon U$ is made block-diagonal, the Lagrangian \eq{eq:action_eff} is unaffected by the matrix elements $\tensor{W}{^m_n}$ with indices $(m,n)$ such that $\tensor{U}{^m_n}=0$. Thus, without loss of generality, we can write
\begin{gather}
L_D =  \sum_{j=1}^J \mathrm{Tr}
				 \int \mathrm{d}^4 x \,  \mathrm{d}^4 p \, [[D_{\rm eff}]]_j\,[[W]]_j, 
\end{gather}
where $[[D_{\rm eff}]]_j \simeq [[\Lambda + \epsilon U]]_j$ and $[[W]]_j$ represent the $j$th matrix block of $\Lambda + \epsilon U$ and $W$, respectively. Hence, nonresonant eigenmodes are decoupled while resonant eigenmodes that belong to the same block remain coupled.

%In this moment, it is important to highlight the differences between the method shown here and the formulations used for studying ray-splitting mode conversion \cite{Friedland:1985ko,Friedland:1987um,Tracy:1993el,Flynn:1994cn,Tracy:2003bl, Tracy:2007jp, Richardson:2008fl}. As explained in \Ref{Friedland:1985ko}, when wave rays traverse regions of mode-conversion, their eigenvectors $\vec{e}_q(x,p)$ can change substantially. In consequence, derivatives of $Q_0(x,p)$ are no longer $\mc{O}(\epsilon)$ in the mode-conversion region, and the assumed ordering in \Eq{eq:G} breaks down. For this reason, mode conversion cannot be described using the procedure shown in this work. However, our procedure is adequate to describe polarization-related phenomena. In particular, note that polarization precession involves interacting eigenmodes that are nearly degenerate in the entire phase space. In this case, the eigenvectors vary smoothly, and the assumed ordering in \Eq{eq:G} is kept. In summary, both mode conversion and polarization precession are wave phenomena that involve the transfer of quanta between different wave modes; however, these processes are fundamentally different, and different methods are needed to describe them.

%%%%%%%%%%%%%%%%%%%%%%%%%%%%%%%%%%%%%%%
%%%%%%%%%%%%%%%%%%%%%%%%%%%%%%%%%%%%%%%
%%%%%%%%%%%%%%%%%%%%%%%%%%%%%%%%%%%%%%%
\section{Reduced action}
\label{sec:reduced}

%%%%%%%%%%%%%%%%%%%%%%%%%%%%%%%%%%%%%%%
\subsection{Basic equations}

Now that blocks of mutually nonresonant modes are decoupled, let us focus on the dynamics of modes within a single block of some size $N$. Hence, the block index will be dropped, and we adopt
\begin{subequations}
	\begin{gather}	
		 L_\tau  \doteq  - (i/2)	
		 					\int \mathrm{d}^4 x  \, 
		 					\left[ \psi^\dag ( \pd_\tau \psi )  - ( \pd_\tau \psi^\dag ) \psi 	\right],
		 \label{eq:L1_new}\\
		 L_D 		\doteq	 \mathrm{Tr} \int \mathrm{d}^4 x \,  \mathrm{d}^4 p \, 
					[[\Lambda + \epsilon U ]] \,  [[W]].
		\label{eq:L2_new}
	\end{gather}
\end{subequations}
Here $\psi$ is a complex-valued function with $N$ components, and $[[W]]$ is the $N \times N$ Wigner tensor with elements
\begin{equation}
	\tensor{[[W]]}{^m_n}(\tau,x,p) = \int \frac{\mathrm{d}^4 s}{(2\pi)^4} \, 
			e^{i  p \cdot s  } \braket{x+ \frac{s}{2} | \psi^m}  \braket{ \psi_n | x- \frac{s}{2} }.
	\label{eq:Wigner_new}
\end{equation}

Since we consider the coupled dynamics of some $N$ resonant modes, only $N$ columns of $Q_0$ actually contribute to $[[D_{\rm eff}]]$. For clarity, let us denote the resonant eigenmodes as $\vec{e}_q$ with indices $q = 1,..., N$. Then, in order to calculate $[[U]]$, one can use \Eq{eq:G}. After block-diagonalizing $U$ and introducing the $\bar{N} \times N$ matrix
\begin{equation}
	\Xi(x,p) = [ \vec{e}_1(x,p), \, ... \,  , \, \vec{e}_N(x,p) ],
	\label{eq:Xi}
\end{equation}
one obtains
\begin{align}
	[[U]] = & \, \frac{i}{4}  \{ \Xi^\dag, \Xi \} \Lambda
								+ \frac{i}{4}  \Lambda \{ \Xi^\dag, \Xi \} 
								+  \frac{i}{2} \{  \Lambda,  \Xi \}_{ \Xi^\dag} \notag \\
				&				+\frac{i}{2} \{ \Xi^\dag  ,   \Lambda \}_{ \Xi } 
								- \frac{i}{2} \{ \Xi^\dag  , \Xi \}_D ,
	\label{eq:U_block}
\end{align}
which is a $N \times N$ Hermitian matrix. 

Furthermore, it is convenient to split $[[D_{\rm eff}]]$ as follows:
\begin{gather}
	[[D_{\rm eff}]] = \lambda \mathbb{I}_N + \epsilon\mc{U},
	\label{eq:Deff_block}
\end{gather}
where $\lambda \doteq N^{-1} \text{Tr}\,[[D_{\rm eff}]]$ is the average of the eigenvalues of $[[D_{\rm eff}]]$ and $\epsilon \mc{U} \doteq [[D_{\rm eff}]] - \lambda \mathbb{I}_{\bar{N}}$ is the remaining traceless part of $[[D_{\rm eff}]]$.

In the special case when all $\lambda^{(q)}$ within the block are identical and $[[U]]$ is traceless, then $\Lambda = \lambda \mathbb{I}_N $, and $\mc{U} = [[U]]$. We call such modes degenerate. Then, the expression \eq{eq:U_block} for $[[U]]$ simplifies, and one obtains
\begin{align}
	\mc{U}(x,p)		
				=		&	\, 	\frac{i}{4}  \{ \Xi^\dag, \Xi \} \lambda
							+ 	\frac{i}{4} \lambda \{ \Xi^\dag, \Xi \} 
							+ 	\frac{i}{2} \{  \lambda,  \Xi \}_{\Xi^\dag} 
							\notag \\
						&	+	\frac{i}{2} \{ \Xi^\dag  ,   \lambda \}_{\Xi} 
							- 	\frac{i}{2} \{ \Xi^\dag  , \Xi \}_D
							\notag \\
				=		&	\, 	-\frac{1}{2i} \Xi^\dag \{  \lambda,  \Xi \}
							-	\frac{1}{2i} \{ \Xi^\dag  ,   \lambda \} \Xi 
							\notag \\
						&	+ 	\frac{1}{2i} \{ \Xi^\dag  , \Xi \}_D
							-  \frac{1}{2i}  \{ \Xi^\dag, \Xi \}_{\lambda }
							\notag \\
				=		&	- 	\left[  \Xi^\dag \{ \lambda,  \Xi \} 	\right]_A
							+ 	\left[ \vphantom{(-\pd_p \lambda)} (\pd_p \Xi^\dag) 
												( D- \lambda \mathbb{I}_N)  (\pd_x \Xi) \right]_A ,
	\label{eq:bar_U_aux}
\end{align}
where we used the bracket introduced in \Eq{eq:poisson2} and the subscript `$A$' denotes ``anti-Hermitian part;'' \ie for any matrix $M$, then $M_A \doteq (M - M^\dag) / (2i)$. The expression in \Eq{eq:bar_U_aux} can also be written more explicitly as
\begin{align}
	\mc{U}(x,p) 	& = \left( -\frac{\pd \lambda}{\pd p_\mu} \right) 
									\left( \Xi^\dag \frac{\pd \Xi}{\pd x^\mu} \right)_A 
							+   \left( \frac{\pd \lambda}{\pd x^\mu} \right) 
									\left( \Xi^\dag \frac{\pd \Xi}{\pd p_\mu} \right)_A 
						\notag \\
						& ~~~	+ 	\bigg[  \frac{\pd \Xi^\dag}{\pd p_\mu}  
										( D- \lambda \mathbb{I}_N ) 
										\frac{\pd \Xi}{\pd x^\mu} \bigg]_A .
	\label{eq:bar_U}
\end{align}

Examples of physical systems, where these simplified formulas are applicable, include spin-$1/2$ particles \cite{Ruiz:2015hq,Ruiz:2015dh} and EM waves propagating in isotropic dielectrics \cite{Ruiz:2015dv}.

%%%%%%%%%%%%%%%%%%%%%%%%%%%%%%%%%%%%%%%
\subsection{Parameterization of the action}
\label{sec:parameterize}

In order to derive the corresponding ELEs, let us adopt the following parameterization:
\begin{equation}
	\psi(\tau,x) = \sqrt{ \mc{I}(\tau,x) } \, z(\tau,x) \, e^{i\theta(\tau,x)}.
	\label{eq:eikonal}
\end{equation}
Here $\theta(\tau,x)$ is a real variable that serves as the rapid phase common for all $N$ modes (remember that all modes within the block of interest are approximately resonant to each other). Also, $\mc{I}(\tau,x)$ is a real function, and $z(\tau,x)$ is a $N$-dimensional complex unit vector ($z^\dag z=1$), whose components describe the amount of quanta in the corresponding modes. (Since we parameterize the $N$-dimensional complex vector $\psi$ by the $N$-dimensional complex vector $z$ plus two independent real functions $\theta$ and $\mc{I}$, not all components of $z$ are truly independent. For an extended discussion, see \Ref{Ruiz:2015hq}.)

After substituting the ansatz \eq{eq:eikonal} into \Eq{eq:L1_new}, the Lagrangian $L_\tau$ is given by
\begin{equation}
	 L_\tau =  \int \mathrm{d}^4 x  \, \mc{I} \left[ 
		 		 		\pd_\tau \theta - (i \epsilon /2) (  z^\dag \pd_\tau z - \cc )  \right].
	\label{eq:L1_eikonal}
\end{equation}
(Here we formally introduce $\epsilon$ to denote that $z$ is a slowly-varying quantity; however, this ordering parameter will be removed later.) Now, we calculate the Wigner tensor \eq{eq:Wigner_new}. Substituting \Eq{eq:eikonal} into \Eq{eq:Wigner_new}, we obtain
\begin{widetext}
\begin{align}
	\tensor{[[W]]}{^m_n} (\tau,x,p)  
		= 	&	\int \frac{\mathrm{d}^4 s}{(2 \pi )^4 } \,
					\sqrt{ \mc{I}(x+s/2,\tau) } z^m(x+s/2,\tau)  
					e^{i\theta(x+s/2,\tau)} 
					\sqrt{ \mc{I}(x-s/2,\tau) } z_n^*(x-s/2,\tau) 
					e^{-i\theta(x-s/2,\tau)}
					e^{i p \cdot s } \notag \\
		=	&	\int \frac{\mathrm{d}^4 s}{(2 \pi )^4 } \,
					\mc{I} (\tau,x) z^m(\tau,x)  z_n^*(\tau,x) 
					e^{ i ( p-k) \cdot s} 
					\notag \\
			&	 + \epsilon 
				 	\int \frac{\mathrm{d}^4 s}{(2 \pi )^4 } \,
				 	\frac{s^\mu}{2} \cdot 
					\left[ \frac{\pd  ( \sqrt{ \mc{I}} \, z^m) }{\pd x^\mu} \sqrt{ \mc{I}} z_n^*  
						- \sqrt{ \mc{I}} z^m \frac{\pd  ( \sqrt{ \mc{I}} \, z_n^*) }{\pd x^\mu}  \right] \, 
					e^{ i ( p-k) \cdot s}
					+ \mc{O}(\epsilon^2) \notag \\
		=	& ~ \mc{I}(\tau,x) \, z^m(\tau,x)  z_n^*(\tau,x)  \, \delta^4( p-k)	
			 - 		\frac{i\epsilon}{2}  
					\frac{\pd  [\delta^4(p- k)] }{\pd p_\mu} 
					\left[ \frac{\pd  ( \sqrt{ \mc{I}} \, z^m) }{\pd x^\mu} \sqrt{ \mc{I}} z_n^*  
						- \sqrt{ \mc{I}} z^m \frac{\pd  ( \sqrt{ \mc{I}} \, z_n^*) }{\pd x^\mu}  \right] 
					+ \mc{O}(\epsilon^2),
	\label{eq:Wigner_eikonal}
\end{align}
where we introduced the four-wavevector $k_\mu(\tau,x) \doteq -\pd_\mu \theta(\tau,x) = ( \omega, - \vec{k})$, which is considered a slow function. [Accordingly, the contravariant representation is $k^\mu(x, \tau) = ( \omega,  \vec{k})$.] Inserting \Eq{eq:Wigner_eikonal} into \Eq{eq:L2_new} and integrating over the momentum coordinate, we obtain
\begin{align}
	 L_D 	=	&  \int \mathrm{d}^4 x \,  \mathrm{d}^4 p \, 
	 					\left( \lambda \tensor{[[W]]}{^m_m}
	 					+ \epsilon \, \tensor{\mc{U}}{^m_n} \tensor{[[W]]}{^n_m} 	\right)
					\notag \\
	 		=	&	 \int \mathrm{d}^4 x \,  \mc{I} \left[   \lambda (x,k)   z^\dag z
	 					+\epsilon  z^\dag \mc{U} (x,k)  z 	\right]
	 				- \frac{i \epsilon}{2}  \int \mathrm{d}^4 x \, \mathrm{d}^4 p \,   
					 \lambda \frac{\pd  [\delta^4(p- k) ]}{\pd p_\mu} 
					 \left[ \frac{\pd  ( \sqrt{ \mc{I}} \, z^m) }{\pd x^\mu} \sqrt{ \mc{I}} z_m^*  
						- \cc \right]  
					  +\mc{O}(\epsilon^2)
					 \notag \\
	 		=	&	  \int \mathrm{d}^4 x \,  \mc{I} \left[ \lambda (x,k) 
	 					+\epsilon  z^\dag \mc{U}(x,k)  z 	\right]
	 				- \frac{i \epsilon}{2}  \int \mathrm{d}^4 x \,  \mc{I} \, 
					v^\mu (\tau, x)
					 \left( z^\dag \frac{\pd z}{\pd x^\mu}  - \cc \right)
					 +\mc{O}(\epsilon^2),
	\label{eq:L2_eikonal}
\end{align}
\end{widetext}
where we integrated by parts and used $z^\dag z =1$. Here 
\begin{equation}
	v^\mu (\tau, x) \doteq 
			- \left[\frac{ \pd \lambda (x,p)}{ \pd p_\mu} \right]_{p=k(\tau, x)}
	\label{eq:GO_vel}
\end{equation}
is the zeroth-order (in $\epsilon$) group velocity of the wave. We then introduce the convective derivative
\begin{equation}
	\frac{\mathrm{d}}{\mathrm{d} \tau} \doteq  \frac{\pd}{\pd \tau }  
			+   v^\mu (\tau, x) \frac{\pd}{\pd x^\mu}.
	\label{eq:convective_d}
\end{equation}
Summing \Eqs{eq:L1_eikonal} and \eq{eq:L2_eikonal}, we obtain the action $\smash{\mc{S} = \int \mathrm{d} \tau \, L + \mc{O}(\epsilon^2)}$, where the Lagrangian is given by
\begin{align}
	L= & \int  \mathrm{d}^4 x  \,  \mc{I} \, \bigg[
			 \pd_\tau \theta + \lambda(x,k)  
			\notag \\
			&
			- \frac{i\epsilon }{2} 
			\left( z^\dag  \frac{\mathrm{d} z}{\mathrm{d} \tau} 
											- \frac{\mathrm{d} z^\dag }{\mathrm{d} \tau} z \right) 
			+ \epsilon z^\dag  \mc{U}(x,k)  z 
			\bigg] .
	\label{eq:lagr_eikonal}
\end{align}

Equation \eq{eq:lagr_eikonal}, along with the definitions in \Eqs{eq:Xi}-\eq{eq:Deff_block}, \eq{eq:GO_vel}, and \eq{eq:convective_d}, is the main result of this work. The first line on the right-hand side of \Eq{eq:lagr_eikonal} represents the lowest-order GO Lagrangian. The terms in the second line of \Eq{eq:lagr_eikonal} are $\mc{O}(\epsilon)$ and introduce polarization effects. [Importantly, diffraction terms would be $\mc{O}(\epsilon^2)$ and thus are safe to neglect in our first-order theory.] In what follows, we discuss the consequences of this theory and provide an example, where we apply the theory to study polarization effects on EM waves in weakly magnetized plasmas.

%%%%%%%%%%%%%%%%%%%%%%%%%%%%%%%%%%%%%%%
%%%%%%%%%%%%%%%%%%%%%%%%%%%%%%%%%%%%%%%
%%%%%%%%%%%%%%%%%%%%%%%%%%%%%%%%%%%%%%%
\section{Traditional geometrical optics}
\label{sec:GO}

%%%%%%%%%%%%%%%%%%%%%%%%%%%%%%%%%%%%%%%
\subsection{Continuous wave model}
\label{sec:GO_wave}

To lowest order in $\epsilon$, the Lagrangian \eq{eq:lagr_eikonal} can be approximated simply with
\begin{equation}
	L_{\rm GO} \doteq \int  \mathrm{d}^4 x ~ \mc{I} \, \left[ \, \pd_\tau \theta + \lambda (x, k) \, \right],
	\label{eq:lagr_GO}
\end{equation}
which one may interpret as a Hayes-type representation \cite{Hayes:1973dt} of the GO wave Lagrangian in the extended $(\tau,x)$ space. This Lagrangian is parameterized by just two functions, the rapid phase $\theta$ and the total action density $\mc{I}$. Thus, varying the action $\mc{S}_{\rm GO} = \int \mathrm{d}\tau \, L_{\rm GO}$, we obtain the following ELEs:
\begin{subequations}	\label{eq:ELE_scalar}
	\begin{align}
		\delta \theta: 	\quad		&		\pd_\tau \mc{I} + \pd_\mu ( v^\mu \mc{I} )= 0,
				\label{eq:act}		\\
		\delta \mc{I}: 	\quad		&		\pd_\tau \theta + \lambda(x, k ) = 0,
				\label{eq:hj}
	\end{align}
\end{subequations}
where  $v^\mu(\tau, x)$ is the GO four-group-velocity \eq{eq:GO_vel}.

As mentioned in \Sec{sec:basic}, the dynamics of the physical wave propagating in spacetime is obtained by adopting $\pd_\tau \Psi=0$, which also corresponds to $\pd_\tau \mc{I} = \pd_\tau \theta =0$. Hence, \Eqs{eq:ELE_scalar} become
\begin{subequations}	\label{eq:ELE_scalar_physical}
	\begin{gather}
		\frac{\pd}{\pd t} \left( - \frac{\pd \lambda}{\pd \omega} \, \mc{I} \right) 
				+ \del \cdot \left( \frac{\pd \lambda}{\pd \vec{k}} \, \mc{I} 	\right) = 0,
				\label{eq:act_physical}		\\
				\lambda(x, k) = 0.
				\label{eq:hj_physical}
	\end{gather}
\end{subequations}
Equation \eq{eq:act_physical} is the action conservation theorem, or the photon conservation theorem. Equation \eq{eq:hj_physical} is the local dispersion relation. For an in-depth discussion of these equations, see, \eg \Refs{Tracy:2014to,Dodin:2012hn}.

%%%%%%%%%%%%%%%%%%%%%%%%%%%%%%%%%%%%%%%
\subsection{Point-particle model}
\label{sec:GO_point}

The ray equations corresponding to the above field equations can be obtained as the point-particle limit. In this limit, $\mc{I}$ can be approximated with a delta function
\begin{equation}
	\mc{I}(\tau,x) = \mc{I}_0 \delta^4 \boldsymbol{(} x-X(\tau) \boldsymbol{)}.
	\label{eq:point}
\end{equation}
Here $\mc{I}_0$ denotes the total action, which is conserved according to \Eq{eq:act_physical}. The value of $\mc{I}_0$ is not essential below so we adopt $\mc{I}_0 = 1$ for brevity. 

In this representation, the wave packet is located at the position $X(\tau)$ in space-time, and the independent parameter is $\tau$. [This means that at a given $\tau$, the wave packet is located at the spatial point $\vec{X}(\tau)$ at time $t(\tau)$.] When inserting \Eq{eq:point} into \Eq{eq:lagr_GO}, the first term in the action gives the following:
\begin{align} 
\int  \mathrm{d} \tau \, \mathrm{d}^4 x & \, \mc{I} \, \pd_\tau \theta \notag \\
		= & \int  \mathrm{d} \tau \, \mathrm{d}^4 x \,
			  \delta^4 \boldsymbol{(} x-X(\tau) \boldsymbol{)} \,\pd_\tau \theta(\tau,x) 
			  \notag\\
		= & -\int \mathrm{d} \tau \, \mathrm{d}^4 x \,  \theta(\tau,x)  
				[ \pd_\tau \delta^4 \boldsymbol{(} x-X(\tau) \boldsymbol{)}] 
				\notag\\
		= & \int  \mathrm{d} \tau \, \mathrm{d}^4 x \, \theta(\tau,x)  
				[ \dot{X}^\mu (\tau) \pd_\mu \delta^4 \boldsymbol{(} x-X(\tau) \boldsymbol{)}]  \notag\\
		= & - \int  \mathrm{d} \tau \, \mathrm{d}^4 x ~
				  \pd_\mu \theta(\tau,x) \dot{X}^\mu (\tau)  
				  \delta^4 \boldsymbol{(} x-X(\tau) \boldsymbol{)}  \notag\\
		= &  \int \mathrm{d} \tau \, P_\mu (\tau) \dot{X}^\mu (\tau) ,
\label{eq:s1del}
\end{align}
where $P_\mu (\tau) \doteq - \pd_\mu \theta\boldsymbol{(} \tau, X(\tau) \boldsymbol{)} $. Similarly,
\begin{equation}
	\int \mathrm{d}^4 x \,  \delta^4 \boldsymbol{(} x-X(\tau) \boldsymbol{)} 
			\lambda(x, -\pd \theta ) =   \lambda\boldsymbol{(} X(\tau), P(\tau)\boldsymbol{)} . 
\end{equation}
Thus, the point-particle action is expressed as
\begin{equation}
	\mc{S}_{\rm GO} = \int \mathrm{d} \tau \, 
		\left[  P (\tau) \cdot \dot{X} (\tau)   + \lambda ( X, P ) \right].
		\label{eq:act_GO}
\end{equation}

This is a covariant action, where $X(\tau)$ and $P(\tau)$ serve as canonical coordinates and canonical momenta, respectively. Treating $X$ and $P$ as independent variables leads to ELEs matching Hamilton's covariant equations
\begin{subequations}	\label{eq:point_GO}
	\begin{align}
		\delta P_\mu: 	\quad		&		
				\frac{ \mathrm{d} X^\mu}{\mathrm{d} \tau} = - \frac{\pd \lambda}{\pd P_\mu} , \\
		\delta X^\mu: 	\quad		&		
				\frac{ \mathrm{d} P_\mu}{\mathrm{d} \tau} = \frac{\pd \lambda}{\pd X^\mu} .
	\end{align}
\end{subequations}
These are the commonly known ray equations; for instance, see \Ref{Tracy:2014to}. They can also be written as
\begin{align*}
		\frac{ \mathrm{d} X^0}{\mathrm{d} \tau} & = - \frac{\pd \lambda}{\pd P_0} , &
		\frac{ \mathrm{d} \vec{X}}{\mathrm{d} \tau} & = \frac{\pd \lambda}{\pd \vec{P}} 	,	
		\\
		\frac{ \mathrm{d} P^0}{\mathrm{d} \tau} & = \frac{\pd \lambda}{\pd X^0} , &
		\frac{ \mathrm{d} \vec{P}}{\mathrm{d} \tau} & = - \frac{\pd \lambda}{\pd \vec{X}} .
\end{align*}

Note that the first term in the integrand in \Eq{eq:act_GO} represents the symplectic part of the canonical phase-space Lagrangian, and the second term represents the Hamiltonian part. Since the Hamiltonian part $\lambda(X,P)$ does not depend explicitly on $\tau$, then $ \mathrm{d} \lambda(X,P) / \mathrm{d} \tau=0$ along the ray trajectories. Thus, the ray dynamics lies on the dispersion manifold defined by
\begin{equation}
	\lambda ( X , P ) = 0.
\end{equation}
As a reminder, $\lambda(x,p)$ is defined as the average eigenvalue of the resonant block, \ie $\lambda \doteq N^{-1} \text{Tr}\,[[D_{\rm eff}]]$. The GO action \eq{eq:act_GO} is only accurate to lowest order in $\epsilon$; hence, one can approximate $\lambda (x,p) \simeq \lambda^{(n)}(x,p)$, where $\lambda^{(n)}$ is any particular resonant eigenvalue. This occurs because the resonant eigenvalues differ by $\mc{O}(\epsilon)$ and because the polarization coupling is also $\mc{O}(\epsilon)$.

%%%%%%%%%%%%%%%%%%%%%%%%%%%%%%%%%%%%%%%
%%%%%%%%%%%%%%%%%%%%%%%%%%%%%%%%%%%%%%%
%%%%%%%%%%%%%%%%%%%%%%%%%%%%%%%%%%%%%%%
\section{Extended geometrical optics}
\label{sec:XGO}

In this section, we explore the polarization effects determined by the Lagrangian \eq{eq:lagr_eikonal}. For the sake of conciseness, we only discuss the point-particle ray dynamics. For an overview of the continuous-wave model, see \Ref{Ruiz:2015hq}.

%%%%%%%%%%%%%%%%%%%%%%%%%%%%%%%%%%%%%%%
\subsection{Point-particle model}
\label{sec:XGO_point}

The ray equations with polarization effects included can be obtained as a point-particle limit of the Lagrangian \eq{eq:lagr_eikonal}. As in \Sec{sec:GO_point}, we approximate the wave packet to a single point in spacetime [\Eq{eq:point}]. As shown in \Refs{Ruiz:2015hq,Ruiz:2015bz}, the Lagrangian \eq{eq:lagr_eikonal} can be replaced by a point-particle Lagrangian so the action is
\begin{align}
	\mc{S}_{\rm XGO}
			= & \int \mathrm{d} \tau \, \bigg[ \, 
				    P \cdot \dot{X} - ( i  /2) ( Z^\dag \dot{Z} - \dot{Z}^\dag Z )  \notag  \\ 
			   & + \lambda(X, P) +  Z^\dag \mc{U}(X, P) Z \,  \bigg] ,
	\label{eq:action_XGO_point}
\end{align}
where $Z(\tau)\doteq z\boldsymbol{(} \tau, X(\tau)  \boldsymbol{)}$ is the point-particle polarization vector and we dropped the GO ordering parameter $\epsilon$. In the complex representation, $Z$ and $Z^\dag$ are canonical conjugate, and
\begin{gather}\label{eq:Z}
	Z^\dag(\tau) Z(\tau) = 1.
\end{gather}
Even though the components of $Z$ are not independent by definition (\Sec{sec:parameterize}), it can be shown \cite{Ruiz:2015hq} that treating them as independent in this point-particle model leads to correct results provided that the initial conditions satisfy \Eq{eq:Z}. Hence, the independent variables in $S_{\rm XGO}$ are $(X, P, Z, Z^\dag)$, and the corresponding ELEs are
\begin{subequations}	\label{eq:XGO_ELEs}
	\begin{align}
			\delta P_\mu: 	\quad		&		
					\frac{ \mathrm{d} X^\mu}{\mathrm{d} \tau} = - \frac{\pd \lambda}{\pd P_\mu} 
					-   Z^\dag  \frac{\pd \mc{U} }{\pd P_\mu} Z, 
					\label{eq:XGO_X}	\\
			\delta X^\mu: 	\quad		&		
					\frac{ \mathrm{d} P_\mu}{\mathrm{d} \tau} = \frac{\pd \lambda}{\pd X^\mu} 
					+    Z^\dag  \frac{\pd \mc{U} }{\pd X^\mu} Z, 
					\label{eq:XGO_P} 	\\
			\delta Z^\dag :		\quad 	&	
					\frac{ \mathrm{d} Z}{\mathrm{d} \tau} = -i \mc{U}	 Z , 
					\label{eq:XGO_Z}	\\	
			\delta Z :				\quad 	&	
					\frac{ \mathrm{d} Z^\dag}{\mathrm{d} \tau} = i Z^\dag \mc{U}
					\label{eq:XGO_Zdag}	 .	
	\end{align}
\end{subequations}
Together with \Eqs{eq:Xi}-\eq{eq:Deff_block}, \Eqs{eq:XGO_ELEs} form a complete set of equations. The first terms on the right-hand side of \Eqs{eq:XGO_X} and \eq{eq:XGO_P} describe the ray dynamics in the GO limit. The second terms describe the coupling to the mode polarization. Equations \eq{eq:XGO_Z} and \eq{eq:XGO_Zdag} describe the wave-polarization dynamics.

As in \Sec {sec:GO_point}, the Hamiltonian part of \Eq{eq:action_XGO_point} is constant along the ray trajectories. As before, the ray dynamics lies on the dispersion manifold defined by setting the Hamiltonian part to zero; \ie
\begin{equation}
	\lambda(X,P) + Z^\dag \mc{U}(X,P) Z =0.
	\label{eq:dispersion_XGO}
\end{equation}
As a reminder, $\lambda(x,p)$ is defined through \Eq{eq:Deff_block} as $\lambda \doteq N^{-1} \text{Tr}\,[[D_{\rm eff}]]$, and $\mc{U} \doteq [[D_{\rm eff}]] - \lambda \mathbb{I}_{\bar{N}}$ is the remaining traceless part of $[[D_{\rm eff}]]$. 

%%%%%%%%%%%%%%%%%%%%%%%%%%%%%%%%%%%%%%%
\subsection{Precession of the wave spin}

Let us also describe the rotation of $Z(\tau)$ as follows. Since $\mc{U}(X,P)$ is a traceless Hermitian $N \times N$ matrix, it can be decomposed into a linear combination of $N^2-1$ generators $\mc{T}_u$ of $\text{SU}( N )$, which are traceless Hermitian matrices, with some real coefficients $-W^u$ \cite{foot:anisovich}:
\begin{gather}
	\mc{U} = - \sum_{u = 1}^{N^2 - 1} 
					\mc{T}_u W^u \equiv - \vec{\mc{T}} \cdot \vec{W} .
\end{gather}
Then, we introduce the $(N^2-1)$-dimensional vector
\begin{equation}
	\vec{S}(\tau) \doteq Z^\dag(\tau) \vec{\mc{T}} Z(\tau)
\end{equation}
so that $Z^\dag \mc{U} Z = - \vec{S} \cdot \vec{W}$. The components of $\vec{S}(\tau)$ satisfy the following equation:
\begin{align}
	\mathrm{d}_\tau  S_w 
		& = Z^\dag \mc{T}_w (\mathrm{d}_\tau Z) 
				+ (\mathrm{d}_\tau Z^\dag) \mc{T}_w Z \notag \\
		& = i Z^\dag  \mc{U} \mc{T}_w  Z - i Z^\dag \mc{T}_w \mc{U}  Z \notag \\
		& = i Z^\dag [ \mc{U} ,  \mc{T}_w] Z \notag \\
		& = -i \,Z^\dag [ \mc{T}_u, \mc{T}_w] Z W^u \notag \\
		& =  f_{uwv} (Z^\dag \mc{T}^v Z) W^u \notag \\
		& = f_{wvu} S^v W^u, 
	\label{eq:aux7}
\end{align}
where $f_{abc}$ are structure constants. They are defined via $[ \mc{T}_a, \mc{T}_b] = i f_{abc}  \mc{T}^c$ so that the structure constants $f_{abc}$ are antisymmetric in all indices \cite{foot:anisovich}.

For example, consider the case when only two waves are resonant. Then, $N^2 - 1 = 3$, $\mc{T}^v$ are the three Pauli matrices divided by two (so $|\vec{S}|^2 = 1/2$), and $f_{wuv}$ is the Levi-Civita symbol, so $f_{wvu} S^v W^u = (\vec{S} \times \vec{W})_w$. For a Dirac electron, which is a special case, such $\vec{S}$ is recognized as the spin vector undergoing the well known precession equation, $\mathrm{d}_\tau \vec{S} = \vec{S} \times \vec{W}$ \cite{Ruiz:2015hq}. In optics, this is an equation for the Stokes vector that was derived earlier to characterize the polarization of transverse EM waves in certain simple media \cite{Bliokh:2008km,Kravtsov:2007cl,Bliokh:2007fr}.

Hence, it is convenient to extend this quantum terminology also to $N$ resonant waves. We will call the corresponding $(N^2 - 1)$-dimensional vector $\vec{S}$ a generalized ``wave-spin" vector and express $f_{wvu} S^v W^u$ symbolically as $(\vec{S} * \vec{W})_w$, where `$*$' can be viewed as a generalized vector product. Notably, using the concept of spin vector $\vec{S}$, one can rewrite \Eqs{eq:XGO_Z} and \eq{eq:XGO_Zdag} as follows:
\begin{gather}\label{eq:spineq}
	\frac{ \mathrm{d} }{\mathrm{d} \tau }  \vec{S} = \vec{S} * \vec{W},
\end{gather}
which is understood as a generalized precession equation.

In the particular case when $\vec{S}$ is conserved (we call such waves ``pure states''), then \Eqs{eq:XGO_X}, \eq{eq:XGO_P}, and \eq{eq:spineq} form a closed set of equations, and $\lambda-\vec{S} \cdot \vec{W}$ serves as an effective scalar Hamiltonian. The dynamics of $Z$ and $Z^\dag$ does not need to be resolved in this case, so one can rewrite $S_{\rm XGO}$ as a functional of $(X,P)$ alone:
\begin{gather}
	S_{\rm XGO} = \int \mathrm{d} \tau \, [ 
		P \cdot \dot{X} + \lambda(X,P) - \vec{S} \cdot \vec{W}(X, P) ].
\end{gather}
An example of the dynamics described by such action will be discussed in \Sec{sec:EM_pure}.

A more general case is when $\vec{S}$ is close to some eigenvector $\vec{w}$ of $\vec{W}$ that corresponds to some nondegenerate  eigenvalue $\Omega_w$. If $\Omega_w$ is large enough, then $\vec{S}(\tau)$ will remain close to $\vec{w}(\tau)$ and will only experience small-amplitude oscillations. These oscillations can be understood as a generalized \textit{zitterbewegung} effect \cite{foot:zitter}, and they are transient, \ie vanish when $\dot{\vec{W}}$ becomes zero. In this regime, no mode conversion occurs at $\tau \to \infty$. In contrast, if $\Omega_w$ is not large enough, the change of $\vec{S}$ governed by \Eq{eq:spineq} is not necessarily negligible. This corresponds to mode conversion and causes ray splitting at $\tau \to \infty$ (see, \eg \Refs{Friedland:1985ko,Friedland:1987um,Tracy:1993el,Flynn:1994cn,Tracy:2003bl, Tracy:2007jp, Richardson:2008fl}). This is discussed below.

%%%%%%%%%%%%%%%%%%%%%%%%%%%%%%%%%%%%%%%
\subsection{Mode conversion as a form of spin precession}

Equation \eq{eq:XGO_Z} [and thus \Eq{eq:spineq}] can also describe mode conversion as it is understood in \Refs{Friedland:1985ko,Friedland:1987um,Tracy:1993el,Flynn:1994cn,Tracy:2003bl, Tracy:2007jp, Richardson:2008fl}. This is shown as follows. Let us consider the resonant interaction between two modes as an example; then, $\mc{U}$ is a $2 \times 2$ matrix. From \Eq{eq:Deff_block}, $\mc{U}$ is Hermitian and traceless and can be represented as
\begin{equation}
	\mc{U}(X,P) = 
	\begin{pmatrix}
		\Delta \lambda	/2 &
		U_{12} \\
		U_{12}^* &
		- \Delta \lambda /2
	\end{pmatrix},
\end{equation}
where $\Delta\lambda (X,P) \doteq [ \lambda^{(1)} - \lambda^{(2)} ]/2 + ( U_{11} - U_{22})/2$ and the coefficient $U_{12}$ determines the mode coupling. Suppose that, absent coupling ($U_{12} = 0$), the dispersion curves of two modes cross at some point $(X_*, P_*)$. Suppose also that $\Delta \lambda (\tau) = \Delta \lambda \boldsymbol{(} X(\tau) , P(\tau) \boldsymbol{)} $ changes along the ray trajectory approximately linearly in $\tau$. Then, $\Delta \lambda \approx \alpha \tau$, where $\alpha$ is some constant coefficient and we chose the origin on the time axis such that $\Delta \lambda(\tau = 0) = 0$ for simplicity. Similarly, $ U_{12} (\tau) \doteq U_{12} \boldsymbol{(} X(\tau) , P(\tau) \boldsymbol{)} \simeq \beta + \gamma \tau$, where $\beta$ and $\gamma$ are some constants. Assuming $\beta$ is sufficiently large, we neglect the term $\gamma \tau$ for it only causes a correction to the dominant effect. Thus, near the mode-conversion region, \Eq{eq:XGO_Z} is approximately written as
\begin{equation}
	i \frac{\mathrm{d}}{\mathrm{d} \tau}
	\begin{pmatrix}
		Z_1 \\ Z_2
	\end{pmatrix}
	=
	\begin{pmatrix}
		\alpha \tau/2 	& \beta 					\\
		\beta^*				&	-\alpha \tau/2
	\end{pmatrix}
	\begin{pmatrix}
		Z_1 \\ Z_2
	\end{pmatrix}.
	\label{eq:Z_mode}
\end{equation}

Equation \eq{eq:Z_mode} is the well-known equation for mode conversion that was studied by Zener in \Ref{Zener:1932iz}. After eliminating $Z_2$, the governing equation for $Z_1$ is
\begin{equation}
	\ddot{Z}_1(\tau) + \left(  |\beta|^2 +i \frac{\alpha}{2} + \frac{\alpha^2 \tau^2}{4} \right) Z_1(\tau)=0.
\end{equation}
Letting $w \doteq \tau \sqrt{\alpha} \, e^{i \pi / 4} $ and $n \doteq - i |\beta|^2 / \alpha$, the equation above can be written as a Weber equation 
\begin{equation}
	Z_1''(w) + \left( n + \frac{1}{2} - \frac{w^2}{4} \right) Z_1(w) =0,
\end{equation}
whose solutions are the parabolic cylinder functions $D_n(w)$. In \Refs{Zener:1932iz,Flynn:1994cn}, the matrix connecting the waves entering and exiting the resonance are obtained by analyzing asymptotics of $D_n(w)$. Specifically,
\begin{equation}
	\begin{pmatrix}
		Z_{\rm 1,out} \\ Z_{\rm 2,out}
	\end{pmatrix}
	=
	\begin{pmatrix}
		\mc{T} 	& - \mc{C}^* \\
		\mc{C}	&	\mc{T}
	\end{pmatrix}
	\begin{pmatrix}
		Z_{\rm 1, in} \\ Z_{\rm 2, in}
	\end{pmatrix},
\end{equation}
where
\begin{gather}
	\mc{T}   	= 	\exp ( - \pi |\eta|^2 ) , \quad \quad	
	\mc{C}  	= 	- \frac{\sqrt{2 \pi \eta}}{ \eta \Gamma( -i |\eta|^2 )},
\end{gather}
where $\Gamma$ is the Gamma function and $\eta \doteq \beta / \sqrt{\alpha}$. The transmission and conversion coefficients for the wave quanta are, correspondingly,
\begin{gather}
	|\mc{T}| = \exp ( -  2 \pi |\beta|^2/ |\alpha| ),\\
	|\mc{C}|^2 = 1- |\mc{T}|^2 .
\end{gather}
(Also see \Ref{Tracy:1993el} for a somewhat different approach leading to the same answer.)

This calculation shows that mode conversion, in the way as commonly described in literature \cite{Friedland:1985ko,Friedland:1987um,Tracy:1993el,Flynn:1994cn,Tracy:2003bl, Tracy:2007jp, Richardson:2008fl}, is nothing but a manifestation of the wave-spin precession described by \Eqs{eq:XGO_Z} and \eq{eq:spineq}. Note that the present point-particle model cannot capture ray-splitting because it introduces only one ray for the whole field. However, this theory does predict the transfer of wave quanta, which is a prerequisite for ray-splitting. For a complete analysis on ray-splitting mode conversion, please refer to \Refs{Tracy:2003bl, Tracy:2007jp, Tracy:2014to}.

%%%%%%%%%%%%%%%%%%%%%%%%%%%%%%%%%%%%%%%
%%%%%%%%%%%%%%%%%%%%%%%%%%%%%%%%%%%%%%%
%%%%%%%%%%%%%%%%%%%%%%%%%%%%%%%%%%%%%%%
\section{Discussion: Waves in weakly magnetized plasmas}
\label{sec:discussion}

A simplified form of the theory above was applied to describe spin-$1/2$ particles \cite{Ruiz:2015hq,Ruiz:2015dh} and waves in isotropic dielectrics \cite{Ruiz:2015dv}. Here we present another example of its application, namely, EM waves in weakly magnetized cold plasmas. (The case of strongly magnetized plasmas will be discussed in a separate paper.) We assume that the plasma response is determined by particles of just one type, \eg electrons. The generalization to multi-component plasma is straightforward to do.

%%%%%%%%%%%%%%%%%%%%%%%%%%%%%%%%%%%%%%%
\subsection{Dispersion operator}

The linearized equations of motion are \cite{Stix:1992waves}
\begin{subequations}	\label{eq:stix}
	\begin{gather}
		\pd_t \vec{v} = 	\, (q/m) \vec{E} + (q/ mc) \vec{v} \times \vec{B}_0, \\
		\pd_t \vec{E} = 	- 4 \pi q n_0 \vec{v} + c \del \times \vec{B}, \\
		\pd_t \vec{B} = 	-c\del \times \vec{E}.
	\end{gather}
\end{subequations}
Here $q$, $m$, $n_0(\vec{x})$, and $\vec{v}(t,\vec{x})$ are the particle charge, mass, unperturbed background density, and flow velocity, respectively. Also, $\vec{E}(t,\vec{x})$ denotes the perturbation electric field, $\vec{B}(t,\vec{x})$ is the perturbation magnetic field, $\vec{B}_0(\vec{x})$ is the background magnetic field, and $c$ is the speed of light. We introduce a re-scaled velocity field $\bar{\vec{v}}(t,\vec{x}) \doteq  \vec{v}(t,\vec{x})  [4\pi n_0 (\vec{x}) m]^{1/2}$, so
\begin{subequations}	\label{eq:stix_normal}
	\begin{gather}
		\pd_t \bar{\vec{v}} = 		\, \omega_p \vec{E} 
									+  \bar{\vec{v}} \times \vec{\Omega} , \label{eq:moment} 	\\
		\pd_t \vec{E} = 			-		\omega_p \bar{\vec{v}} + c \del \times \vec{B}, \\
		\pd_t \vec{B} = 	-c \del \times \vec{E} 				\label{eq:faraday}, 
	\end{gather}
\end{subequations}
where $\omega_p(\vec{x})	\doteq [4\pi q^2 n_0 (\vec{x}) /m]^{1/2} $ is the plasma frequency and $\vec{\Omega}(\vec{x})\doteq q\vec{B}_0(\vec{x})/(mc) $ is the gyrofrequency.

Let us write \Eqs{eq:stix_normal} using the abstract Hilbert space notation. Let $\ket{\vec{v} } $ be a state vector representing the velocity field such that $\vec{v}(x) = \braket{ x | \vec{v} }$. Likewise, we introduce $\ket{\vec{E} }$ and $\ket{\vec{B}}$ as the state vectors of $\vec{E}(x)$ and $\vec{B}(x)$, respectively. Then, \Eqs{eq:stix_normal} can be written as follows:
\begin{subequations}	\label{eq:stix_Hilbert}
\begin{gather}
		\hat{p}_0 \ket{ \bar{\vec{v}}  } 
								= 	i \hat{\omega}_p \ket{ \vec{E} } 
									- (\vec{\alpha} \cdot \hat{\vec{\Omega}}  ) \ket{ \bar{\vec{v}}  } , 
								\label{eq:v_abstract} \\
		\hat{p}_0 \ket{ \vec{E} }  
								= 	- i \hat{\omega}_p  \ket{ \bar{\vec{v}}  } 
									+ i c (\vec{\alpha} \cdot \hat{\vec{p}}) \ket{ \vec{B} }, 
								\label{eq:E_abstract} \\
		\hat{p}_0 \ket{ \vec{B} } 
								= -i c (\vec{\alpha} \cdot \hat{\vec{p}}) \ket{ \vec{E} } , 
								\label{eq:B_abstract}
	\end{gather}
\end{subequations}
where $\hat{\omega}_p \doteq \omega_p(\hat{\vec{x}})$ and $\hat{\vec{\Omega}} \doteq \vec{\Omega}(\hat{\vec{x}})$. (As a reminder, $\smash{ \hat{p}_0= i \pd_t}$ and $\smash{ \hat{\vec{p}}= - i \del }$ are the components of the four-momentum operator in the $x$-representation.) Also, $\vec{\alpha} \doteq \left( \alpha^1, \alpha^2, \alpha^3 \right)$ are $3\times 3$ Hermitian matrices \cite{foot:GellMann}
\begin{subequations}\label{eq:alpha}
	\begin{align} 
\alpha^1 \doteq & 
	\begin{pmatrix}
		0 & 0 & 0 \\
		0 & 0 & -i \\
		0 & i & 0 
	\end{pmatrix}, \\ 
\alpha^2 \doteq &
	\begin{pmatrix}
		0 & 0 & i \\
		0 & 0 & 0 \\
		-i & 0 & 0 
	\end{pmatrix}, \\ 
\alpha^3 \doteq	&
	\begin{pmatrix}
		0 & -i & 0 \\
		i & 0 & 0 \\
		0 & 0 & 0 
	\end{pmatrix}.
	\end{align}			
\end{subequations}
These matrices serve as generators for the vector product. Namely, for any two column vectors $\vec{A}$ and $\vec{B}$, one has
\begin{subequations} \label{eq:property}
	\begin{gather}
 		( \vec{\alpha} \cdot \vec{A}) \vec{B}			= i \vec{A} \times \vec{B}, 	
 				\label{eq:property1}	 	\\
 		\vec{A}^{\rm T} \alpha^j \vec{B}			= -i (\vec{A} \times \vec{B})^j, 
 				\label{eq:property2}				 	
	\end{gather}	
\end{subequations}
where the superscript `T' denotes the matrix transpose.

The next step is to construct a dispersion operator for the electric field state $\ket{\vec{E}}$. Starting from \Eq{eq:v_abstract}, we solve for the velocity field in terms of the electric field. Hence, we formally obtain the following:
\begin{align}
	\ket{\bar{\vec{v}} } = 
					& 	\, i \hat{\omega}_p 
						( \hat{p}_0 \mathbb{I}_3+ \vec{\alpha}\cdot \hat{\vec{\Omega}}  )^{-1} 
						\ket{\vec{E} }
						\notag \\
				=	& 	\, i  \hat{\omega}_p  \left[ \frac{1}{\hat{p}_0} 
						- \frac{\vec{\alpha}\cdot  \hat{\vec{\Omega}} }
								{\hat{p}_0^2 - \hat{\Omega}^2}
						+ \frac{(\vec{\alpha}\cdot \hat{\vec{\Omega}} )^2}
								{\hat{p}_0(\hat{p}_0^2 - \hat{\Omega}^2)} \right] 
						\ket{\vec{E} },
\end{align}
where $ \hat{\Omega} \doteq |\vec{\Omega}(\hat{\vec{x}})|$. Similarly, we obtain $\smash{\ket{ \vec{B} } = -i c (\vec{\alpha} \cdot \hat{\vec{p}}) \hat{p}_0^{-1} \ket{ \vec{E} } }$ from \Eq{eq:B_abstract}. Substituting these results into \Eq{eq:E_abstract}, we obtain
\begin{equation}
	\oper{D}  \ket{\vec{E} } =0,
\end{equation}
where
\begin{equation}
	\oper{D}  \doteq
			-\hat{p}_0^2 
			+ (\vec{\alpha}\cdot \hat{\vec{p}} )^2 
			+ \hat{\omega}_p^2 
			- \frac{\hat{\omega}_p^2 \hat{p}_0( \vec{\alpha}\cdot \hat{\vec{\Omega}} )} 		
					{\hat{p}_0^2 - \hat{\Omega}^2} 
			+ \frac{ \hat{\omega}_p^2 (\vec{\alpha}\cdot \hat{\vec{\Omega}} )^2}
						{   \hat{p}_0^2 - \hat{\Omega}^2 }
	\label{eq:E_oper}
\end{equation}
serves as the dispersion operator for $\ket{ \vec{E} }$. (For convenience, we let $c=1$.) Since $\omega_p(\vec{x})$ and $\vec{\Omega}(\vec{x})$ are independent of time, then $\smash{\hat{p}_0}$ commutes with $\smash{ \hat{\omega}_p}$ and $\smash{ \hat{\vec{\Omega}} }$, so $ \smash{\oper{D}  }$ is manifestly Hermitian. The corresponding action \eq{def:action} for the electric field is $\smash{ \mc{S}  = \braket{ \vec{E}  | \oper{D}  | \vec{E} } }$, and the extended action \eq{eq:action} is 
\begin{equation}
	\mc{S}_{\rm X}  \doteq  \int \mathrm{d}\tau \, \left[
					-\frac{i}{2}	\left(  \braket{ \vec{E} | \pd_\tau   \vec{E} } - \cc \right)
							+ \braket{ \vec{E} | \oper{D} | \vec{E} } \right].
\end{equation}
Note that $\vec{E}$ is a three-dimensional vector field, so $\bar{N} = 3$.

%%%%%%%%%%%%%%%%%%%%%%%%%%%%%%%%%%%%%%%
\subsection{EM waves in weakly magnetized plasma}

We now follow the procedure given in Secs. \ref{sec:eigen} and \ref{sec:reduced} to block-diagonalize the dispersion operator. The Weyl symbol of $\oper{D}$ is
\begin{equation}
	D  \doteq
			-p_0^2 
			+ (\vec{\alpha}\cdot \vec{p} )^2 
			+ \omega_p^2 
			- \frac{ \omega_p^2 p_0( \vec{\alpha}\cdot \vec{\Omega}  )  } 		
					{ p_0^2 - \Omega^2} 
			+ \frac{ \omega_p^2 (\vec{\alpha}\cdot \vec{\Omega} )^2}
						{   p_0^2 - \Omega^2 }.
	\label{eq:disp_symbol}
\end{equation}
For the sake of simplicity, we consider the case of a wave propagating in a weakly magnetized plasma. (The general case will be described in a separate paper.) Thus, supposing that the typical wave frequency is much larger than the gyrofrequency $(\omega \sim p_0 \gg \Omega )$, we expand the dispersion symbol \eq{eq:disp_symbol} in powers of $\Omega/ p_0$:
\begin{equation}
	D  \simeq D_0 + D_1 + \mc{O}( \Omega^2/p_0^2),
\end{equation}
where
\begin{subequations}
	\begin{gather}
			D_0 (\vec{x},p)  		 =  	-p_0^2 + (\vec{\alpha}\cdot \vec{p} )^2 
												+ \omega_p^2(\vec{x}), \\
			D_1 (\vec{x},p_0)		 = 	-\omega_p^2(\vec{x}) 
												( \vec{\alpha}\cdot \vec{\Omega} ) / p_0 .
	\end{gather}
\end{subequations}
To simplify the following calculation, we assume that $D_1 \sim \mc{O}(\Omega/ p_0)$ is comparable in magnitude to the GO parameter $\epsilon$, but this is not essential. Hence, we will consider $D_1$ as a perturbation only. 

Following \Sec{sec:eigenmode}, the next step is to identify the eigenvalues and eigenmodes of the dispersion symbol $D_0 (\vec{x},p)  $. The corresponding eigenvalues are
\begin{subequations}
	\begin{gather}
		\lambda^{(1)} (\vec{x},p)	= 		\, -p\cdot p  + \omega_p^2 (\vec{x}), \\
		\lambda^{(2)} (\vec{x},p)	= 		\, -p\cdot p  + \omega_p^2 (\vec{x}), \\
		\lambda^{(3)} (\vec{x},p_0)	= 		\, -p_0^2 + \omega_p^2 (\vec{x}),
	\end{gather}
\end{subequations}
where $p \cdot p = p_0^2 - \vec{p}^2$. These eigenvalues correspond to the dispersion relations of two transverse EM waves and of longitudinal Langmuir oscillations, respectively. The matrix $Q_0$ defined in \Eq{eq:T0} is given by
\begin{equation}
	Q_0(\vec{p}) = 	[	\, \vec{e}_1(\vec{p}), \,
								\vec{e}_2	(\vec{p}), \,
								\vec{e}_\vec{p}(\vec{p}) \,
							],
\end{equation}
where $\vec{e}_1(\vec{p})$ and $\vec{e}_2(\vec{p})$ are any two orthonormal vectors in the plane normal to $\vec{e}_\vec{p}(\vec{p}) \doteq \vec{p}/ |\vec{p}|$. A right-hand convention is adopted such that $\smash{\vec{e}_1 \times \vec{e}_2 = \vec{e}_\vec{p}}$. One can easily verify that these vectors are indeed eigenvectors of $D_0(\vec{x},p)$. For example,
\begin{align}
	D_0 \, \vec{e}_1 	&	=	[-p_0^2 
											+ (\vec{\alpha}\cdot \vec{p} )(\vec{\alpha}\cdot \vec{p} ) 
											+ \omega_p^2] \,
				 						\vec{e}_1
				 						\notag \\
								&	=	(-p_0^2  + \omega_p^2) \, \vec{e}_1
										- \vec{p}  \times ( \vec{p} \times \vec{e}_1 )
				 						\notag \\
								&	=	(-p_0^2  + \vec{p}^2 + \omega_p^2) \, \vec{e}_1
										\notag \\
								&	=	\lambda^{(1)}  \, \vec{e}_1 ,
\end{align}
where \Eq{eq:property1} was used. Similar calculations follow for the other two eigenmodes $\vec{e}_2$ and $\vec{e}_\vec{p}$.

We now analyze the dynamics of the transverse EM waves. From \Sec{sec:reduced}, the eigenvalue is $\smash{ \lambda (\vec{x},p) = - p \cdot p + \omega_p^2 (\vec{x})}$, and $\smash{ \Xi(\vec{p}) = [ \vec{e}_1(\vec{p}) , \, \vec{e}_2(\vec{p}) \, ]}$ is a $3\times 2$ matrix. Since $\Xi(\vec{p})$ only depends on the spatial momentum coordinate, then the polarization-coupling Hamiltonian $\mc{U}$ in \Eq{eq:bar_U} is given by
\begin{align}
	\mc{U}(\vec{x},\vec{p}) 	&	=	\frac{\pd \lambda}{\pd x^\mu}
											\left( \Xi^\dag  \frac{\pd \Xi}{\pd p_\mu}\right)_A
											\notag \\
								&	=	-\frac{\pd \lambda}{\pd \vec{x}} \cdot 
											\left( \Xi^\dag  \frac{\pd \Xi}{\pd \vec{p} }\right)_A
											\notag \\
								&	=	 -\frac{\pd \lambda}{\pd \vec{x}} \cdot
										\left[ 
											\begin{pmatrix}
												\vec{e}^1 \\
												\vec{e}^2		
											\end{pmatrix}
											\frac{\pd }{\pd \vec{p} }
											\begin{pmatrix}
												\vec{e}_1 &
												\vec{e}_2		
											\end{pmatrix}
											\right]_A
										\notag \\
							 	&	=	-\frac{\pd \lambda}{\pd \vec{x}} \cdot
											\begin{pmatrix}
												\vec{e}^1 	\frac{\pd }{\pd \vec{p}} \vec{e}_1	&
												 \vec{e}^1 	\frac{\pd }{\pd \vec{p} } \vec{e}_2 	\\
												\vec{e}^2 	\frac{\pd }{\pd \vec{p}} \vec{e}_1	&
												\vec{e}^2	\frac{\pd }{\pd \vec{p}} \vec{e}_2		
											\end{pmatrix}_A
										\notag \\
							 	&	=	-\frac{1}{i}
							 				\frac{\pd \lambda}{\pd \vec{x}} \cdot 
											\begin{pmatrix}
												0	&
												 \vec{e}^1 	\frac{\pd }{\pd \vec{p}} \vec{e}_2 	\\
												-\vec{e}^1	\frac{\pd }{\pd \vec{p}} \vec{e}_2	 &
												0	
											\end{pmatrix},
	\label{eq:U_aux}
\end{align}
where $\vec{e}^q$ is the dual to $\vec{e}_q$, so $\vec{e}^q \vec{e}_r= \delta_r^q$. (Specifically, $\smash{ \vec{e}^q }$ is a row vector, whose elements are complex-conjugate of those of $\vec{e}_q	$.) Since $\del \lambda =  \del 
\omega_p^2$, we can also write \Eq{eq:U_aux} in the form
\begin{equation}
	\mc{U} (\vec{x},\vec{p}) =  - \sigma_y(\del \omega_p^2) \cdot \vec{F} ,
	\label{eq:EM_U}
\end{equation}
where $\sigma_y$ is the $y$-component of the Pauli matrices 
\begin{equation}
	\sigma_y = 	\begin{pmatrix}
							0		&		-i		\\
							i		&		0	
						\end{pmatrix}
\end{equation}
and $\vec{F}(\vec{p})$ is a vector with components given by
\begin{equation}
	\vec{F} (\vec{p}) \doteq \vec{e}^1 	\frac{\pd }{\pd \vec{p}} \vec{e}_2 .
	\label{eq:Fdef}
\end{equation}
For example, one may choose
\begin{align}
	\vec{e}_1(\vec{p}) \doteq 	
		\begin{pmatrix}  
				\frac{p_x p_z }{p \sqrt{ p_x^2+p_y^2 } } \\ 
				\frac{p_y p_z}{ p \sqrt{ p_x^2+p_y^2 } } \\ 
				- \frac{ \sqrt{ p_x^2+p_y^2 } }{ p } \end{pmatrix} , 
	& &
	\vec{e}_2(\vec{p}) \doteq 	
		\begin{pmatrix}  
				-\frac{p_y }{  \sqrt{ p_x^2+p_y^2 } } \\ 
				\frac{p_x }{  \sqrt{ p_x^2+p_y^2 } } \\ 
				0 
		\end{pmatrix},
	\label{eq:evec}
\end{align}
so that
\begin{equation}
	\vec{F} (\vec{p}) = \frac{ \vec{p}_\perp \times \vec{p} }{ |\vec{p} | | \vec{p}_\perp |^2  },
\end{equation}
where $\vec{p}_\perp \doteq \begin{pmatrix}	p_x, p_y, 0 \end{pmatrix}^T$; or, more explicitly,
\begin{equation}
	\vec{F} (\vec{p}) = \frac{p_z}{   |\vec{p} | | \vec{p}_\perp |^2 }
		\begin{pmatrix}
				 p_y		\\ 			
				-p_x 		\\ 		
				0 
		\end{pmatrix}.		
	\label{eq:def_F}
\end{equation}
(The specific choice of $\vec{e}_1$ and $\vec{e}_2$ does not affect the resulting equations within the accuracy of the present theory. For more details, see \Sec{sec:noncanonical}.)

\begin{figure*}
	\includegraphics[scale=.55]{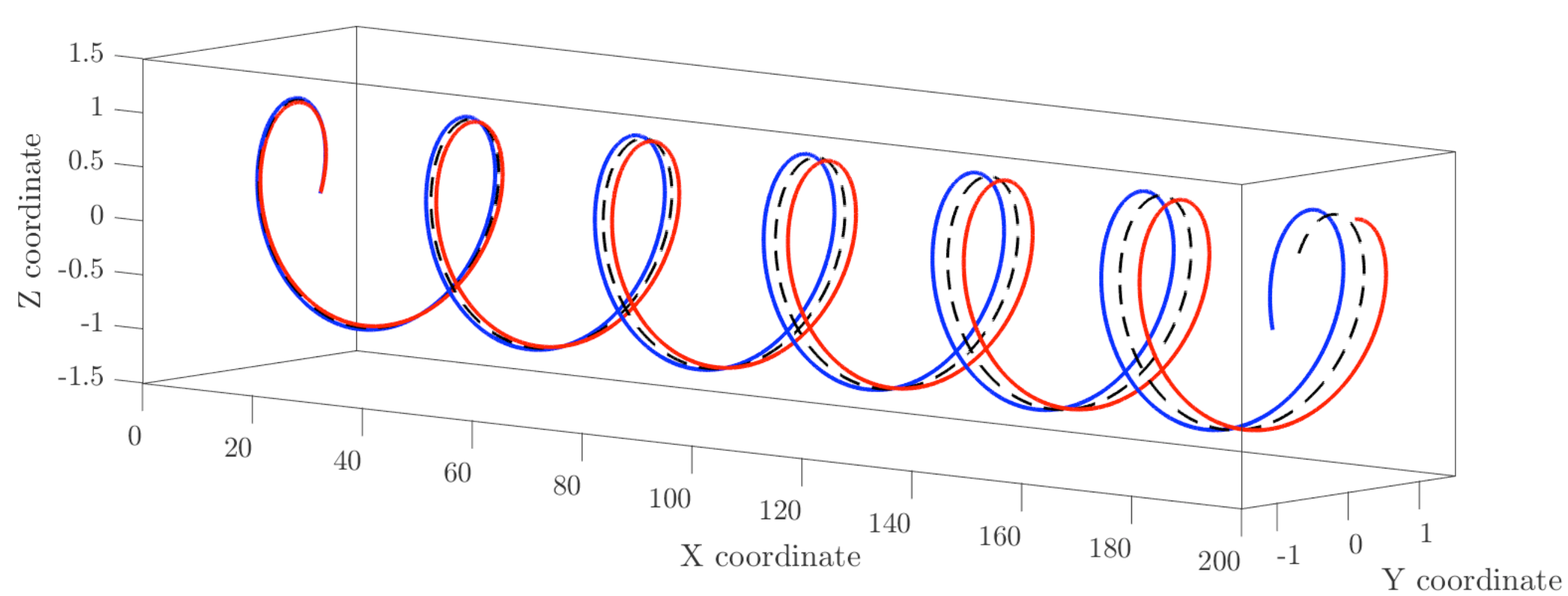}
	\caption{Comparison between ray trajectories calculated using the equations of traditional GO [\Eqs{eq:EM_ELE_GO}, dashed line] and extended GO [\Eqs{eq:ELE_pure}]. The blue and red lines represent the ray trajectories for the right-hand and left-hand polarized rays, respectively. For simplicity, nonmanetized plasma is considered, so the Faraday effect is absent. The plasma frequency is given by  $\omega_p^2(\vec{x}) = y^2 + z^2$. The initial location of the ray trajectories is $\vec{X}_0=(0,1,0)$, and the initial momentum is $\vec{P}_0=(5,0,1)$. (The units are arbitrary, since the figure is a general illustration only.) For this simulation, the GO parameter is roughly $\epsilon \sim 1 / ~ |\vec{P}_0| \sim 0.2$. Due to the radial gradient in the plasma frequency, the wave rays follow helical trajectories along the $x$ axis.}
	\label{fig:polarization}
\end{figure*}

Returning to the perturbation caused by the background magnetic field, the projection of the eigenmodes on the matrix $D_1(\vec{x},p_0)$  is given by
\begin{align}
	\Xi^\dag D_1 \Xi 
		&	=	-\frac{\omega_p^2}{p_0}
				\begin{pmatrix}
						\vec{e}^1 \\
						\vec{e}^2  	
				\end{pmatrix} 
				( \vec{\alpha}\cdot \vec{\Omega})
				\begin{pmatrix}
						\vec{e}_1 &
						\vec{e}_2		
				\end{pmatrix} 
				\notag \\	
		&	=	-\frac{\omega_p^2}{p_0}
				\begin{pmatrix}
						\vec{e}^1  ( \vec{\alpha}\cdot \vec{\Omega})\vec{e}_1&
						\vec{e}^1  ( \vec{\alpha}\cdot \vec{\Omega})\vec{e}_2	\\
						\vec{e}^2  ( \vec{\alpha}\cdot \vec{\Omega})\vec{e}_1 &
						\vec{e}^2  ( \vec{\alpha}\cdot \vec{\Omega})\vec{e}_2		
				\end{pmatrix} 
				\notag \\	
		&	=	i \frac{\omega_p^2}{p_0}
				\begin{pmatrix}
						0	&
						\vec{e}_\vec{p} \cdot \vec{\Omega}		\\
						-\vec{e}_\vec{p} \cdot \vec{\Omega} &
						0					
				\end{pmatrix} 
				\notag \\		
		&	=	-\frac{\omega_p^2}{p_0} (\vec{e}_\vec{p} \cdot \vec{\Omega}) \sigma_y ,
		\label{eq:EM_perturbation}
\end{align}
where we used \Eq{eq:property2}.

%%%%%%%%%%%%%%%%%%%%%%%%%%%%%%%%%%%%%%%
\subsection{Ray dynamics}

Now, let us discuss the point-particle ray dynamics. Following \Sec{sec:XGO}, we substitute $\lambda (\vec{x},p) = - p \cdot p + \omega_p^2 (\vec{x})$, \Eq{eq:EM_U}, and \Eq{eq:EM_perturbation} into \Eq{eq:action_XGO_point}. We then obtain the point-particle action
\begin{multline}
	\mc{S}_{\rm XGO} = \int \mathrm{d} \tau \, \big[ \, 
				P \cdot \dot{X}  -  (i/2) ( Z^\dag \dot{Z} - \dot{Z}^\dag Z )  \\
			 	-  P \cdot P 
			 	+ \omega_p^2(\vec{X})  
			 	+ \Sigma (\vec{X}, P) Z^\dag \sigma_y  Z \,  \big] ,
	\label{eq:EM_act_point}
\end{multline}
where the polarization-coupling matrix is given by
\begin{equation}
	\Sigma (\vec{x}, p) \doteq 
			- (\del \omega_p^2) \cdot \vec{F}
			- \frac{\omega_p^2}{p_0} (\vec{e}_\vec{p} \cdot \vec{\Omega})
	\label{eq:Sigma}
\end{equation}
and $Z(\tau)$ is a complex-valued vector with two components that describe the degree of polarization along the vectors $\vec{e}_1$ and $\vec{e}_2$. It is normalized such that $Z^\dag Z =1$. 

In the action \eq{eq:EM_act_point}, the two polarization modes are coupled through the Pauli matrix $\sigma_y$. However, these modes can be decoupled when using the basis of circularly polarized modes. We introduce the variable transformation
\begin{equation}
	Z(\tau)= \, \mc{R} \, \Gamma(\tau),
	\label{eq:Z_transformation}
\end{equation}
where
\begin{equation}
	\mc{R} \doteq \frac{1}{\sqrt{2}}
			\begin{pmatrix}
				1	&  1	\\		
				i	&	-i
			\end{pmatrix} 		
\end{equation}
and $\Gamma(\tau)$ is a new vector with components denoted as
\begin{equation}
	\Gamma(\tau)  \doteq 			
				\begin{pmatrix}
					\Gamma_+ \\
					\Gamma_-
				\end{pmatrix}.
\end{equation}
Inserting \Eq{eq:Z_transformation} into the action \eq{eq:EM_act_point} leads to
\begin{multline}
	\mc{S}_{\rm XGO} 	=   \int \mathrm{d} \tau \, \left[ \, 
				P \cdot \dot{X}   
					-  (i/2) ( \Gamma^\dag \dot{\Gamma} - \dot{\Gamma}^\dag \Gamma ) \right. \\
			 	 \left. -  P \cdot P 
			 	 + \omega_p^2(\vec{X})  
			 	+  \Sigma (\vec{X}, P) \Gamma^\dag \sigma_z  \Gamma \,  \right] ,
	\label{eq:EM_act_point_II}
\end{multline}
where $\sigma_z$ is another Pauli matrix,
\begin{equation}
\sigma_z=
	\begin{pmatrix}
		1		&		0	\\
		0		&		-1		
	\end{pmatrix}.
\end{equation}
Here $\Gamma_\pm(\tau) $ represent the wave quanta belonging to the right-hand and left-hand circularly polarized modes, respectively (as defined from the point of view of the source). Also, $\Gamma$ is normalized such that $\Gamma^\dag  \Gamma =1$.

Treating $X(\tau)$, $P(\tau)$, $\Gamma(\tau)$, and $\Gamma^\dag(\tau)$ as independent variables, we obtain the following ELEs:
\begin{subequations}	\label{eq:EM_ELE}
	\begin{align}
		\delta P_\mu: & 	\quad  	\frac{\mathrm{d} X^\mu}{\mathrm{d} \tau} 
											= 2	P^\mu 
												-  \frac{\pd  \Sigma }{\pd P_\mu}  \Gamma \sigma_z \Gamma , 
												\label{eq:EM_ELE_a} \\
		 \delta X^\mu: & 	\quad  	\frac{\mathrm{d} P_\mu}{\mathrm{d} \tau} 
											= \frac{\pd \omega_p^2 }{\pd X^\mu} 
											+ \frac{\pd  \Sigma }{\pd X^\mu}  \Gamma \sigma_z \Gamma , 
											\label{eq:EM_ELE_b} \\
		\delta \Gamma^\dag: & 	\quad  	\frac{\mathrm{d} \Gamma}{\mathrm{d} \tau} 
											=  	-i \Sigma \sigma_z \Gamma ,
											\label{eq:EM_ELE_Gamma} \\
		\delta \Gamma: 		  & 	\quad  	\frac{\mathrm{d} \Gamma^\dag}{\mathrm{d} \tau} 
											=  	i \Gamma \Sigma \sigma_z  .							
	\end{align}
\end{subequations}
Together with \Eq{eq:Sigma}, \Eqs{eq:EM_ELE} form a complete set of equations. The first terms on the right-hand side of \Eqs{eq:EM_ELE_a} and \eq{eq:EM_ELE_b} describe the ray dynamics in the GO limit. The second terms describe the coupling of the mode polarization and the ray curvature.

%%%%%%%%%%%%%%%%%%%%%%%%%%%%%%%%%%%%%%%%%%%%%%%%
\subsection{Restating the Faraday effect}
\label{sec:faraday}

To better understand the polarization equations, let us rewrite \Eq{eq:EM_ELE_Gamma} as an equation in the basis of linearly polarized modes:
\begin{gather}
	\dot{Z} 	= \mc{R} \dot{\Gamma} 
				= -i \Sigma \mc{R} \sigma_z  \Gamma 
				= -i\Sigma (\mc{R} \sigma_z \mc{R}^{-1}) Z 
				= -i\Sigma \sigma_y Z.
\end{gather}
[This equation could also be obtained if the ray equations were derived directly from the action \eq{eq:EM_act_point}.] Since $\Sigma$ is a scalar and $\sigma_y$ is constant, this can be readily integrated, yielding \cite{foot:exponent}
\begin{gather}
	Z(\tau) = \exp( -i \Theta \sigma_y) Z_0 = (\mathbb{I}_2 \cos\Theta - i \sigma_y \sin \Theta) Z_0,
\end{gather}
where $\Theta(\tau) \doteq \int^\tau_0 \mathrm{d} \tau' \, \Sigma(\vec{X}(\tau'), P(\tau'))$ is the polarization precession angle and $Z_0 \doteq Z(\tau=0)$. This result can be also be expressed explicitly as follows:
\begin{gather}
	Z(\tau) = \left(
		\begin{array}{cc}
			\cos\Theta & -\sin\Theta \\
			\sin\Theta & \cos\Theta \\
		\end{array}
		\right) Z_0.
\end{gather}
It is seen that the polarization of the EM field rotates at the rate $\Sigma(t)$ in the reference frame defined by the basis vectors $(\vec{e}_1, \vec{e}_2)$. The first term in \Eq{eq:Sigma} is identified as the rate of change of the wave Berry phase \cite{Berry:1984jv}. (In optics, the rotation of the polarization plane caused by the Berry phase is also known as the \textit{Rytov rotation} \cite{foot:RytovVladimirskii,Tomita:1986jo,Bliokh:2007fr}.) The second term in \Eq{eq:Sigma} is identified as the rate of change due to Faraday rotation.

%%%%%%%%%%%%%%%%%%%%%%%%%%%%%%%%%%%%%%%
\subsection{Dynamics of pure states}
\label{sec:EM_pure}

If a ray corresponds to a strictly circular polarization such that $\sigma_z \Gamma = \pm \Gamma$, the action \eq{eq:EM_act_point_II} can be simplified to $ \mc{S}_{\rm XGO}  = \int \mathrm{d} \tau \, L_\pm$, where the Lagrangian is given by
\begin{align}
	L_\pm
			= P \cdot \dot{X} -  P \cdot P + \omega_p^2(\vec{X}) 
			 	\pm   \Sigma (\vec{X}, P)  .
		\label{eq:lagr_pure}
\end{align}
Here the Lagrangian $L_\pm$ governs the propagation of right-hand and left-hand polarization modes, respectively. The corresponding ELEs are
\begin{subequations}	\label{eq:ELE_pure}
	\begin{align}
		\delta P_\mu: & 	\quad  	\frac{\mathrm{d} X^\mu}{\mathrm{d} \tau} 
											=  2	P^\mu 
												\mp \frac{\pd \Sigma}{\pd P_\mu} , \\
		 \delta X^\mu: & 	\quad  	\frac{\mathrm{d} P_\mu}{\mathrm{d} \tau} 
											=  \frac{\pd \omega_p^2 }{\pd X^\mu} 
												\pm \frac{\pd  \Sigma }{\pd X^\mu},
	\end{align}
\end{subequations}
or in terms of spacetime components,
\begin{align*}
	\frac{\mathrm{d} X^0}{\mathrm{d} \tau}			
								&		=  2	P^0 
											\mp \frac{ \pd \Sigma}{\pd P_0}, &
	\frac{\mathrm{d} \vec{X} }{\mathrm{d} \tau}	
								&		= 2 \vec{P} \pm \frac{ \pd \Sigma}{\pd \vec{P}} , \\
	\frac{\mathrm{d} P_0}{\mathrm{d} \tau}		
								&		= 0, &
 	\frac{\mathrm{d} \vec{P} }{\mathrm{d} \tau}	
 								&		= -\frac{ \pd \omega_p^2 }{\pd \vec{X}} 
										\mp \frac{ \pd \Sigma }{\pd \vec{X}}   .
\end{align*}
The first terms on the right-hand side of \Eqs{eq:ELE_pure} describe the ray dynamcis in the GO limit. The second terms describe the coupling of the mode polarization and the ray curvature. They are also responsible for the polarization-driven bending of ray trajectories.

As shown, $P_0$ remains constant because the background medium is time independent. In order to obtain the value of $P_0$, we note that the ray Hamiltonian
\begin{equation}
	H_\pm(X,P) =  -P \cdot P + \omega_p^2(\vec{X}) \pm   \Sigma (\vec{X}, P)
	\label{eq:EM_Hamiltonian} 
\end{equation}
is independent of $\tau$, so one can readily verify that
\begin{equation}
	H_\pm \boldsymbol{(} X(\tau) , P(\tau)  \boldsymbol{)}
			=\mathrm{constant}.
	\label{eq:Ham_constant}
\end{equation}
Setting the Hamiltonian equal to zero, we use \Eq{eq:EM_Hamiltonian} to determine $P_0$. One finds 
\begin{equation}
	P_0 \simeq \omega(\vec{X},\vec{P}) 
						\pm \Sigma(\vec{X},P_*) / [ 2 \omega(\vec{X}, \vec{P}) ],
\end{equation}
where $\omega(\vec{X},\vec{P}) \doteq ( \vec{P}^2 + \omega_p^2 )^{1/2} $ is the wave frequency in the GO limit and $P_*^\mu ( \vec{X}, \vec{P} ) \doteq \boldsymbol{(} \omega( \vec{X}, \vec{P} ), \vec{P} \boldsymbol{)} $.

%%%%%%%%%%%%%%%%%%%%%%%%%%%%%%%%%%%%%%%
\subsection{Numerical simulations}

To illustrate the polarization-driven divergence of the ray trajectories, \Fig{fig:polarization} shows the ray trajectories for a right-polarized and left-polarized waves using the Lagrangian \eq{eq:lagr_pure}. For completeness, we also show the calculated ray trajectory as determined by the lowest-order GO ray Lagrangian
\begin{equation}
	L_{\rm GO} = P \cdot \dot{X} -  P \cdot P + \omega_p^2(\vec{X}) ,
	\label{eq:action_spinless}
\end{equation}
which does not account for polarization effects. As anticipated, the ray trajectories predicted by the Lagrangian \eq{eq:lagr_pure} differ noticeably from the ``spinless" ray trajectory predicted by \Eq{eq:action_spinless}; namely;
\begin{subequations}	\label{eq:EM_ELE_GO}
	\begin{align}
		\delta P_\mu: & 	
			\quad  	\frac{\mathrm{d} X^\mu}{\mathrm{d} \tau} 
								=  2	P^\mu  , \\
		 \delta X^\mu: & 	
		 	\quad  	\frac{\mathrm{d} P_\mu}{\mathrm{d} \tau} 
								= \frac{\pd \omega_p^2 }{\pd X^\mu} .
	\end{align}
\end{subequations}
This divergence along the $x$-axis is driven by polarization effects. For EM waves propagating in isotropic non-birefringent dielectrics, this effect is called the Hall effect of light in the optics literature \cite{Bliokh:2008km}.

%%%%%%%%%%%%%%%%%%%%%%%%%%%%%%%%%%%%%%%
\subsection{Noncanonical representation and\\ the Berry connection}
\label{sec:noncanonical}

It is possible to obtain an alternative, noncanonical representation of the ray Lagrangian \eq{eq:lagr_pure} that is invariant with respect to the choice of $\vec{F}(\vec{p})$ for pure states and explicitly shows the so-called Berry connection. Starting from \Eq{eq:lagr_pure} and substituting \Eq{eq:Sigma}, we can write
\begin{align}
	L_\pm
	& = P \cdot \dot{X} -  P \cdot P + \omega_p^2(\vec{X}) \notag \\
	&		\quad	\mp  (\del \omega_p^2) \cdot \vec{F}  
			 			\mp  \frac{\omega_p^2(\vec{X}) }{P_0} 
			 			[\vec{e}_\vec{p} (\vec{P}) \cdot \vec{\Omega}(\vec{X}) ]
		\notag \\
	& \simeq P \cdot \dot{X} -  P \cdot P + \omega_p^2(\vec{X} \mp \vec{F}) 
				\mp  \frac{\omega_p^2(\vec{X}) }{P_0} 
			 			[\vec{e}_\vec{p} (\vec{P}) \cdot \vec{\Omega}(\vec{X}) ]		,
	\label{eq:lagr_noncanonical_aux}
\end{align}
where we assumed that $\omega_p^2(\vec{x})$ is smooth and neglected terms of $\mc{O}(\epsilon^2)$ as usual. Introducing the variables 
\begin{equation}
	x^\mu(\tau) \doteq \boldsymbol{(} X^0, \vec{X} \mp \vec{F}(\vec{P}) \boldsymbol{)} ,
	\quad 
	p_\mu(\tau)  \doteq P_\mu
\end{equation}
and substituting them into \Eq{eq:lagr_noncanonical_aux}, we obtain 
\begin{align}
	L_\pm
	& \simeq p \cdot \dot{x} -  p \cdot p + \omega_p^2(\vec{x}) \notag \\
	&		\quad	\mp \vec{p} \cdot \dot{\vec{F}}
			 			\mp  \frac{\omega_p^2(\vec{x}) }{p_0} 
			 			[\vec{e}_\vec{p} (\vec{p}) \cdot \vec{\Omega}(\vec{x}) ]
		\notag \\
	& = p \cdot \dot{x} -  p \cdot p + \omega_p^2(\vec{x}) \notag \\
	&		\quad	\pm \dot{\vec{p}} \cdot \vec{F}(\vec{p})
			 			\mp  \frac{\omega_p^2(\vec{x}) }{p_0} 
			 			[\vec{e}_\vec{p} (\vec{p}) \cdot \vec{\Omega}(\vec{x}) ]	,
	\label{eq:lagr_noncanonical}
\end{align}
where we dropped a perfect time derivative. We also approximated $\vec{X} \simeq \vec{x}$ in the Faraday rotation term in \Eq{eq:lagr_noncanonical} since it is already $\mc{O}(\epsilon)$. Note that $| \vec{x} -\vec{X} |$ is of the order of the wavelength, \ie small enough to make $\vec{x}$ and $\vec{X}$ equally physical as measures of the ray location.

The term $\dot{\vec{p}} \cdot \vec{F}(\vec{p})$ is known as the Berry connection term \cite{Bliokh:2015bw}. It is to be noted that adding $\pd_\vec{p} \chi(\vec{p}) $ to $\vec{F}(\vec{p})$, where $\chi(\vec{p})$ is an arbitrary scalar function, changes $L_\pm$ by a perfect derivative and does not affect the equations of motion. The ELEs corresponding to the Lagrangian \eq{eq:lagr_noncanonical} are given by
\begin{subequations}
\begin{align}
	\frac{\mathrm{d} x^0}{\mathrm{d} \tau}			
	&		= 2	p_0 
				\mp  \frac{\omega_p^2}{ p_0^2 }
				(\vec{e}_\vec{p} \cdot \vec{\Omega}), \\
	\frac{\mathrm{d} \vec{x} }{\mathrm{d} \tau}	
	&		= 2 \vec{p} 
			\pm \dot{\vec{p}} \times (\del_\vec{p} \times \vec{F} )
			 \mp  \frac{\omega_p^2}{ p_0 } \frac{ \pd  }{\pd \vec{p}} 
				(\vec{e}_\vec{p} \cdot \vec{\Omega})  , 
			\label{eq:EM_ELE_noncanonical}	\\					
	\frac{\mathrm{d} p_0}{\mathrm{d} \tau}		
	&		= 0, \\
 	\frac{\mathrm{d} \vec{p} }{\mathrm{d} \tau}	
 	&		= - \frac{ \pd \omega_p^2}{\pd \vec{x} } 
 				 \pm \frac{ \pd  }{\pd \vec{x}} \left[ \frac{\omega_p^2}{ p_0 }
				(\vec{e}_\vec{p} \cdot \vec{\Omega}) \right]   .
\end{align}
\end{subequations}
These equations are equivalent to \Eqs{eq:ELE_pure} within the accuracy of the theory. Substituting \Eq{eq:def_F}, we can also write \Eq{eq:EM_ELE_noncanonical} as
\begin{equation}
	\frac{\mathrm{d} \vec{x} }{\mathrm{d} \tau}	
	= 2 \vec{p} 
			\pm \frac{ \dot{\vec{p}} \times \vec{p}}{|\vec{p}|^3}
			 \mp  \frac{\omega_p^2}{ p_0 } \frac{ \pd  }{\pd \vec{p}} 
				(\vec{e}_\vec{p} \cdot \vec{\Omega}) .
\end{equation}
Hence, with the use of the noncanonical coordinates $(x,p)$, the equations of motion no longer depend on the specific choice of $\vec{F}(\vec{p})$; \ie they are invariant with respect to the choice \eq{eq:evec} of vectors $\vec{e}_1$ and $\vec{e}_2$. Note that the same equations could be obtained directly from the point-particle limit of \Eq{eq:U_aux}, if one substitutes $-\del \lambda = \dot{\vec{p}}$. For an extended discussion of pure states governed by noncanonical Lagrangians, see \Ref{Littlejohn:1991jv}.

%%%%%%%%%%%%%%%%%%%%%%%%%%%%%%%%%%%%%%%
%%%%%%%%%%%%%%%%%%%%%%%%%%%%%%%%%%%%%%%
%%%%%%%%%%%%%%%%%%%%%%%%%%%%%%%%%%%%%%%
\section{Conclusions}
\label{sec:conclusions}

Even when neglecting diffraction, the well-known equations of geometrical optics (GO) are not entirely accurate. Traditional GO treats wave rays as classical particles, which are completely described by their position and momentum coordinates. However, vector waves have another degree of freedom, namely, their polarization. Polarization dynamics are manifested in two forms: (i) mode conversion, which is the transfer of wave quanta between resonant eigenmodes and can be understood as the precession of the wave spin, and (ii) polarization-driven bending of ray trajectories, which refers to deviations of the GO ray trajectories arising from first-order corrections to the GO dispersion relation. They are easily understood by drawing parallels with quantum mechanics, where similar effects (yet involving $\hbar$) are known as spin rotation and spin-orbital coupling.

In this work, we propose a first-principle variational formulation that captures both types of polarization-related effects simultaneously. We consider general linear nondissipative waves, whose dynamics are determined by some dispersion operator $\smash{\oper{D}}$. Using the Feynman reparameterization and the Weyl calculus, we obtain a reduced Lagrangian model for such general waves. In contrast with the traditional GO Lagrangian, which is $\mc{O}(\epsilon^0)$-accurate in the GO parameter $\epsilon$, our Lagrangian is $\mc{O}(\epsilon)$-accurate. In our procedure, polarization effects are contained in the $\mc{O}(\epsilon)$ corrections to the GO Lagrangian. These corrections may be especially significant for modeling RF waves in laboratory plasmas because such waves can have not-too-small $\epsilon$ (as opposed, for instance, to quantum particles whose spin effects are typically weak). As an example, we apply the formulation to study the polarization-driven divergence of RF waves propagating in weakly magnetized plasma. Assessing the importance of polarization effects on waves propagating in strongly magnetized plasma will be discussed in a separate paper. Likewise, the method of including dissipation \cite{Dodin:2016ut} in the above theory will also be described separately.

This work was supported by the U.S. DOE through Contract No. DE-AC02-09CH11466, by the NNSA SSAA Program through DOE Research Grant No. DE-NA0002948, and by the U.S. DOD NDSEG Fellowship through Contract No. 32-CFR-168a.

\appendix

%%%%%%%%%%%%%%%%%%%%%%%%%%%%%%%%%%%%%%%%%%%%%%
%%%%%%%%%%%%%%%%%%%%%%%%%%%%%%%%%%%%%%%%%%%%%%
\section{The Weyl transform}
\label{app:Weyl}

This appendix summarizes our conventions for the Weyl transform. (For more information, see the excellent reviews in \Refs{Imre:1967fr,McDonald:1988dp,BakerJr:1958bo,Tracy:2014to}.) The Weyl symbol $A(x,p)$ of any given operator $\oper{A}$ is defined as
\begin{equation}
	A(x,p) \doteq 
				\int \mathrm{d}^4 s \, e^{i p \cdot s }
				\braket{ x+ s/2 | \oper{A}	| x	-	s/2	}  ,
	\label{def:weyl_symbol}
\end{equation}
where $p \cdot s = p_0 s_0 - \vec{p}\cdot \vec{s}$ and the integrals span over $\mathbb{R}^4$. We shall refer to this description of the operators as a \textit{phase-space representation} since the symbols \eq{def:weyl_symbol} are functions of the eight-dimensional phase space. Conversely, the inverse Weyl transformation is given by
\begin{equation}
	\oper{A} = \int	
					\frac{\mathrm{d}^4 x \,  \mathrm{d}^4 p \,  	\mathrm{d}^4 s }{(2\pi )^4} \,
					e^{i p \cdot s  / \epsilon } 
					A(x,p) \ket{x-s/2} \bra{x+s/2}.
	\label{eq:weyl_inverse}
\end{equation}
Hence, $\mcu{A}(x,x') = \braket{x|\oper{A}|x'}$ can be expressed as
\begin{equation}
	\mcu{A}(x,x') =  \int
 					\frac{\mathrm{d}^4 p}{(2\pi )^4} \,
					e^{i p \cdot (x'-x)  }
					A		\left(	\frac{x+x'}{2},	p	\right).
	\label{eq:weyl_x_rep}
\end{equation}

In the following, we outline a number of useful properties of the Weyl transform.

\begin{itemize}[leftmargin=*]
\item For any operator $\oper{A}$, the trace $\mathrm{Tr}[\oper{A}] \doteq \int \mathrm{d}^4 x \, \braket{ x | \oper{A} | x }$ can be expressed as
\begin{equation}
	\mathrm{Tr}[\oper{A}] 
				=  \int \frac{\mathrm{d}^4 x \,  \mathrm{d}^4 p }{(2\pi  )^4} \,  
				A(x, p ).
	\label{eq:trace}
\end{equation}

\item If $A(x,p)$ is the Weyl symbol of $\oper{A}$, then $A^\dag(x,p)$ is the Weyl symbol of $\oper{A}^\dag$. As a corollary, the Weyl symbol of a Hermitian operator is real.

\item For any $\oper{C} =\oper{A} \oper{B}$, the corresponding Weyl symbols satisfy \cite{Moyal:1949gj,Groenewold:1946kp}
\begin{equation}
	C (x,p) = A (x,p) \star B (x,p).
	\label{eq:Moyal}
\end{equation}
Here `$\star$' refers to the \textit{Moyal product}, which is given by
\begin{equation}
		A(x,p) \star B (x,p) 
		\doteq 
		A (x,p) e^{i \hat{\mc{L}} /2 }  B (x,p),
	\label{def:Moyal}
\end{equation}
and $\hat{\mc{L}}$ is the \textit{Janus operator}
\begin{equation}
	\hat{\mc{L}}  \doteq  
			\overleftarrow{\pd_p} \cdot \overrightarrow{\pd_x} -		
			\overleftarrow{\pd_x} \cdot \overrightarrow{\pd_p}
			=
			\{ {\cdot} , {\cdot} \}.
\end{equation}
The arrows indicate the direction in which the derivatives act, and $A \hat{\mc{L}} B = \{ A, B \}$ is the canonical Poisson bracket in the eight-dimensional phase space, namely,
\begin{equation}
	\hat{\mc{L}}  =
			\frac{\overleftarrow{\pd} }{\pd p^0}  \frac{ \overrightarrow{\pd} }{\pd x^0}
	-		\frac{\overleftarrow{\pd} }{\pd x^0}  \frac{ \overrightarrow{\pd} }{\pd p^0}
	+		\frac{\overleftarrow{\pd} }{\pd \vec{x}}  \cdot \frac{ \overrightarrow{\pd} }{\pd \vec{p} }
	-		\frac{\overleftarrow{\pd} }{\pd \vec{p}}  \cdot \frac{ \overrightarrow{\pd} }{\pd \vec{x} }.
\end{equation}
Provided that $A\hat{\mc{L}}B$ is small, one can use the following asymptotic expansion of the Moyal product:
\begin{equation}
	A  \star B   \simeq A  \, B  + \frac{i}{2}   \{ A, B \}.
	\label{eq:Moyal_exp}
\end{equation}

\item The Moyal product is associative; \ie 
\begin{equation}
	A \star B \star C 
			= (A \star B) \star C 
			= A \star (B \star C).
\end{equation}

\item Now we tabulate some Weyl transforms of various operators. (We use a two-sided arrow to show the correspondence with the Weyl transform.) First of all, the Weyl transforms of the identity, position, and momentum operators are given by
\begin{equation}
	\hat{1}	\,	\Leftrightarrow	\,	1, \quad
	\hat{x}^\mu \,	\Leftrightarrow	\, x^\mu, \quad
	\hat{p}_\mu \,	\Leftrightarrow	\, p_\mu.
\end{equation}
For any two functions $f$ and $g$, one has
\begin{equation}
	f(\hat{x}) 	\,	\Leftrightarrow	\, f(x), \quad
	g(\hat{p}) 	\,	\Leftrightarrow	\, g(p).
\end{equation}
Similarly, using \Eq{def:Moyal}, one has
\begin{gather}
	\hat{p}_\mu	f(\hat{x}) 	\,	\Leftrightarrow	\, p_\mu f(x)	+ (i/2) \pd_\mu 	f(x), \\
	f(\hat{x}) \hat{p}_\mu	\,	\Leftrightarrow	\, p_\mu f(x)	- (i/2) \pd_\mu 	f(x).
\end{gather}

\end{itemize}

%%%%%%%%%%%%%%%%%%%%%%%%%%%%%%%%%%%%%%%%%


\begin{thebibliography}{49}
\bibitem[{foo({\natexlab{b}})}]{foot:momentum}
\bibinfo{note}{In the $x$-representation, we have the following: $\braket{ x |
  \hat{p}_\mu | \Psi } = \int \mathrm{d}^4 x' \, \braket{ x | \hat{p}_\mu | x'
  } \braket{x' | \Psi} = i \int \mathrm{d}^4 x'\, \pd [ \delta^4(x-x')] / \pd
  x^\mu \Psi(x')= i \pd_\mu \int \mathrm{d}^4 x' \, \delta^4(x-x') \Psi(x') = i
  \pd_\mu \Psi(x)$.}
\bibitem[{foo({\natexlab{e}})}]{foot:GellMann}
\bibinfo{note}{It is to be noted that the $\alpha$ matrices are related to the
  Gell-Mann matrices, which serve as infinitesimal generators of the special
  unitary group $\mathrm{SU}(3)$.}
\bibitem[{foo({\natexlab{f}})}]{foot:exponent}
\bibinfo{note}{Here we used the well known Euler formula for Pauli matrices,
  $e^{i a (\vec{n}\cdot \vec{\sigma}) }=\mathbb{I}_2 \cos a+ i (\vec{n}\cdot
  \vec{\sigma}) \sin a$.}
\end{thebibliography}
\end{document}